\documentclass[apj,numberedappendix]{emulateapj}
\usepackage{epstopdf}
\usepackage{wrapfig}
\usepackage{natbib}
\bibpunct{(}{)}{;}{a}{}{,}

\usepackage{color}
\definecolor{darkgreen}{RGB}{0,142,128}
\definecolor{darkblue}{RGB}{0,100,170}

\begin{document}

\title{The effect of magnetic topology on thermally-driven wind: towards a general formulation of the braking law}

\author{Victor R\'eville$^1$,
        Allan Sacha Brun$^1$,
        Sean P. Matt$^2$, Antoine Strugarek$^{3,1}$, Rui F. Pinto$^{4,1}$}

\affil{$^1$Laboratoire AIM, DSM/IRFU/SAp, CEA Saclay, 91191 Gif-sur-Yvette Cedex, 
France; victor.reville@cea.fr, sacha.brun@cea.fr\\
$^2$ Department of Physics and Astronomy, University of Exeter, Stocker Road, Exeter EX4 4SB, UK; s.matt@exeter.ac.uk\\
$^3$ D\'epartement de physique, Universit\'e de Montr\'eal, C.P. 6128 Succ. Centre-Ville, Montr\'eal, QC H3C-3J7, Canada; strugarek@astro.umontreal.ca\\
$^4$ LESIA, Observatoire de Paris-Meudon, 5 place Jules Janssen, 92195 MEUDON Cedex, FRANCE; rui.pinto@obspm.fr}

\begin{abstract}

Stellar winds are thought to be the main process responsible for the spin down of main-sequence stars. The extraction of angular momentum by a magnetized wind has been studied for decades, leading to several formulations for the resulting torque. However, previous studies generally consider simple dipole or split monopole stellar magnetic topologies. Here we consider in addition to a dipolar stellar magnetic field, both quadrupolar and octupolar configurations, while also varying the rotation rate and the magnetic field strength. 60 simulations made with a 2.5D, cylindrical and axisymmetric set-up and computed with the PLUTO code were used to find torque formulations for each topology. We further succeed to give a unique law that fits the data for every topology by formulating the torque in terms of the amount of open magnetic flux in the wind. We also show that our formulation can be applied to even more realistic magnetic topologies, with examples of the Sun in its minimum and maximum phase as observed at the Wilcox Solar Observatory, and of a young K-star (TYC-0486-4943-1) whose topology has been obtained by Zeeman-Doppler Imaging (ZDI).

\end{abstract}


\section{Introduction} 
\label{sec_intro}

The evolution of the angular momentum of a solar-like star is the result of interactions with its environment: a disk when it's an accreting young star or its expanding atmosphere -the stellar wind- during the main sequence (MS). Despite a broad distribution of rotation rates in very young stars clusters \citep{IrwinBouvier2009}, main-sequence stars spin down and are observed to approximately follow the empirical Skumanich's law: $\Omega(t) \propto t^{-1/2}$ \citep{Skumanich1972}. This law has been deduced from observations of the Pleiades, Ursa Major and the Hyades. Since then observations of low-mass stars open clusters have confirmed this trend. The braking is the consequence of the magnetized wind which carries angular momentum away from the star \citep{Schatzman1959,Schatzman1962} and close planets could play a role too \citep{Strugarek2014}. Also the magnetic activity of the star is stronger with higher rotation rate \citep{Noyes1984,SB2000,Brun2004,Garcia2014} so that more rapidly rotating stars brake more than slower ones. Thus this convergence can be understood, and models have been developed \citep{ReinersMohanty2012,GalletBouvier2013,Brown2014} with a goal of being able to infer age from the stellar rotation period \citep[gyrochronology, see][]{Barnes2003} or magnetic activity \citep[magnetochronology, see][]{Vidotto2014b}.
 
For sun-like stars the wind accelerates mostly due to the pressure gradient \citep{Parker1958} (for a more precise description see \citet{Cranmer2007}). In the case of rapid rotators both magnetic pressure and centrifugal force add to acceleration \citep{WeberDavis1967}. The wind reaches the Alfv\'en speed at the Alfv\'en surface. \citet{WeberDavis1967} demonstrated that for a one dimensional magnetized wind the loss of angular momentum is proportional to the Alfv\'en radius squared, which acts as a lever-arm. Hence models that have been proposed to explain rotation rate evolution need to tie the Alfv\'en radius to the parameters of the problem. \citet{Kawaler1988}, following the formulation of \citet{Mestel1984}, introduced a power-law dependence of the Alfv\'en radius on the strength of the magnetic field over the mass loss rate. This power law formulation has been investigated further by \citet{MP2008} and \citet{Matt2012} who included the influence of the rotation rate.

In those torque formulations the mass-loss rate is assumed to be known, and analytical techniques have been proposed to compute it from stellar parameters \citep{CranmerSaar2011}, but it can also be observed \citep{WoodRev2004}. Indeed the wind is eventually stopped by the interstellar medium pressure and becomes subsonic again, at the termination shock. Beyond, a contact surface between the stellar wind and the interstellar plasma, the astropause, contains heated hydrogen that produces H I Lyman-$\alpha$ absorption, detectable in the UV.  Those data can be used to infer the mass loss rate of sun-like stars \citep{Wood2002} and thus braking models can be constrained by observations and provide a solid base for gyrochronology.

However, to date, studies have mainly considered simple magnetic topologies such as split monopoles \citep{WeberDavis1967,Kawaler1988} and dipoles \citep{Mestel1968,WashShib1993,MP2008,Matt2012,Cohen2014}. The real topology of stellar magnetic field can be much more complex. For instance during the 22 year solar cycle, the sun oscillates between dipolar and quadrupolar dominant topology \citep{DeRosa2012}. It is now generally agreed that magnetic activity in solar-like stars owes its origin to a nonlinear dynamo process operating in and at the base of their convective envelope \citep{Moffatt1978,Brun2004,Charbonneau2010,Brun2013}. High performance numerical simulations are now able to reproduce key characteristics of stellar magnetic activity, such as global scale organization of the magnetic field, regular cycles and flux concentration and emergence \citep{Ghizaru2010,Brown2011,Racine2011,Nelson2013,Kapyla2013,Beaudoin2013}. These simulations inform us on the organization of the large scale magnetic topology in solar-like stars and hence on its impact on their coronal field.

The large scale coronal magnetic fields influence the wind driving and may be responsible for the changes in velocities and mass loss rate over the solar cycle \citep{PintoBrun2011}. Thus models are needed to quantify this effect on the wind driving and the associated extraction of angular momentum. Three dimensional simulations of stellar winds evaluating the mass and angular momentum loss rates have already been made, for instance in the work of \citet{Cohen2014} who used the BATS-R-US code with a dipolar topology and an axisymmetric set-up. \citet{Vidotto2014} introduced realistic topologies of six M-dwarfs, where they provide physical based relations between output and parameters of the simulations. However the sample of stars they used a wide range of stellar parameters whose influence could not be clearly isolated. We chose to work in 2.5D, which means that we have two spatial dimensions, assumed axisymmetry, and vectors have all three components, to be able to perform more than 60 wind simulations in a systematic parametric study. We derive braking laws from our simulations results, which accurately include the influence of magnetic topologies more complex than a dipole, in order to improve rotation evolution models.

In our study we focused on thermally driven winds which are thought to exist in every cool stars with an outer convective envelope (corresponding to spectral type from M to F, \textit{i.e.} 0.1 to 1.4 $M_{\odot}$). We chose to run simulations with a set of 20 initial parameters used in \citet{Matt2012} with different topologies. The first three axisymmetric components that can be extracted from a realistic geometry, the dipole, the quadrupole and the octupole were used for each case. Over this parametric study we have been able to find three braking laws, one for each topology. The range for the parameters cover four orders of magnitude for the rotation rate and two for the magnetic field strength. To go further we propose here a topology-independent formulation for the braking law using the open magnetic flux. It has been preferred to other quantities that appear in the literature, such as the filling factor \citep{CranmerSaar2011}, because it seems to be the simplest -and yet understandable- way to get a topology-independent braking law for all our cases. Moreover we demonstrated that this formulation can be applied for realistic magnetic topology, we show here the example of a young sun whose magnetic field is obtained through ZDI Maps and of the Sun at its minimum and maximum of cycle 22 using Wilcox magnetograms.

We introduce the formulation we use for the stellar wind braking in section \ref{StellarWindModels}, with a more detailed derivation in Appendix \ref{AppA}. Our numerical methods are detailed in section \ref{Numerics}, while in section \ref{Parstudy} we give the results of our parameter study. Section \ref{Collapse} introduce our topology-independent formulation and makes a comparison between the different braking laws. We show then that magnetic torques of realistic topologies can be predicted by our formulation in section \ref{Realistic}. Finally we discuss several points raised by our study in section \ref{discuss} and then come to conclusion and perspective for further work. Appendices \ref{AppB} and \ref{AppC} are dedicated to numerical details of the simulations and to our exhaustive results.

\section{Models for stellar wind braking}
\label{StellarWindModels}

The idea of a non-hydrostatic expanding stellar atmosphere appeared with \citet{Parker1958}, where an accelerated outflow was used to explain the observed pressure ratio of order $10^{14}$ between the base of the corona and the interstellar medium. \citet{Schatzman1962}, \citet{Parker1963}, \citet{WeberDavis1967} and \citet{Mestel1968} then introduced the effect of both  magnetic field and rotation, thus creating the magnetic rotator theory, which is now the standard MHD theory for stellar winds. It combines the driving of the wind due to the pressure gradient and the magneto-centrifugal effect. In ideal MHD, in which the plasma is ``frozen-in'' with the magnetic field, the outflow is directed along the magnetic field lines. Thus the  plasma is rotating with the magnetic field of the star and is subject to a centrifugal force which contribute to the acceleration. Rotation also creates a toroidal magnetic field component ($B_{\phi}$) whose gradient adds also to acceleration through the Lorentz force. \citet{WeberDavis1967} used a simple one dimensional model (at the equator) to quantify the angular momentum carried by the plasma and demonstrated that

\begin{equation}
l \equiv \Omega_* R_A^2,
\end{equation}

where $l$ is the specific (per unit mass) angular momentum of the gas, $\Omega_*$ the rotation rate of the star and $R_A$ the Alfv\'en radius, \textit{i.e.} the radius at which the velocity field reaches the Alfv\'en speed:

\begin{equation}
v_A \equiv \frac{B_p}{\sqrt{4 \pi \rho}},
\end{equation}

where $B_p$ is the poloidal component of the magnetic field in this model. The loss rate of angular momentum by the star is expressed in a steady state as the product of the mass loss rate and the specific angular momentum carried by the outflow, which gives the following expression for the torque exerted on the star:

\begin{equation}
\tau_w = \dot{M}_w \Omega_* R_A^2,
\label{torque}
\end{equation}

where $\dot{M}_w$ is the integrated mass loss rate.

In order to find a formulation for a realistic multi-dimensional outflow, we define a average value for the Alfv\'en radius (which is the cylindrical radius, the distance from the rotation axis) such that equation (\ref{torque}) is always true: 

\begin{equation}
\langle R_A \rangle  = \sqrt{\frac{\tau_w}{\dot{M}_w \Omega_*}}.
\label{AlfRad}
\end{equation}

$\dot{M}_w$ and $\tau_w$ are computed from our simulations once they have reached a steady state. We use in this work the formulation introduced in \citet{Matt2012} (see Appendix \ref{AppA} for a complete derivation of this formulation):

\begin{equation}
\frac{\langle R_A \rangle}{R_*}=K_1 [\frac{\Upsilon}{(1+f^2/K_2^2)^{1/2}}]^m.
\label{form1}
\end{equation}

where 

\begin{equation}
\Upsilon \equiv \frac{B_*^2 R_*^2}{\dot{M}_w v_{esc}}
\label{Ups_def}
\end{equation}

is the magnetization parameter introduced in \citet{MP2008}. A similar parameter has been introduced before in \citet{UdDoula2002}), where the terminal velocity $v_{\infty}$ was used instead of the escape velocity $v_{esc} \equiv \sqrt{(2 G M_*)/R_*}$. Both characterize the magnetization of the wind, which is the ratio of the magnetic field energy and the kinetic energy of the wind. 
 
$f$ is the fraction of break-up rate, \textit{i.e.} the ratio between the rotation rate at the equator of the star (in our simulations the star has a solid body rotation) and the keplerian speed that is defined by:

\begin{equation}
f \equiv \Omega_* R_*^{3/2}(GM_*)^{-1/2}.
\end{equation}

\section{Numerical Setup}
\label{Numerics}

In this work we use the compressible magneto-hydrodynamic (MHD) code PLUTO \citep{Mignone2007}. All simulations are performed in 2.5D (two spatial dimensions, three vector components), assuming axisymmetry and using cylindrical coordinates (hereafter $(R,\phi,Z)$). Since PLUTO is a multi-physics, multi-solver code, we chose a finite-volume method using an approximate Riemann Solver, here the HLL solver \citep{Einfeldt1988}. PLUTO uses a reconstruct-solve-average approach using a set of primitive variables $(\rho, \mathbf{v}, p, \mathbf{B})$ to solve the Riemann problem corresponding to the following conservative ideal MHD equations (expressed here with the set of conservative variables $(\rho, \mathbf{m}, E, \mathbf{B})$): 

\begin{equation}
\label{MHD_1}
\frac{\partial}{\partial t} \rho + \nabla \cdot \rho \mathbf{v} = 0 
\end{equation}
\begin{equation}
\label{MHD_2}
\frac{\partial}{\partial t} \mathbf{m} + \nabla \cdot (\mathbf{mv}-\mathbf{BB}+\mathbf{I}p) = \rho \mathbf{a}
\end{equation}
\begin{equation}
\label{MHD_3}
\frac{\partial}{\partial t} E + \nabla \cdot ((E+p)\mathbf{v}-\mathbf{B}(\mathbf{v} \cdot \mathbf{B})) = \mathbf{m} \cdot \mathbf{a}
\end{equation}
\begin{equation}
\label{MHD_4}
\frac{\partial}{\partial t} \mathbf{B} + \nabla \cdot (\mathbf{vB}-\mathbf{Bv})=0
\end{equation}

where $\rho$ is the density, $p$ and $\mathbf{v}$ are the pressure and the velocity field, $\mathbf{m}=\rho\mathbf{v}$ is the momentum density, $E=\rho \epsilon+\mathbf{m}^2/(2\rho)+\mathbf{B}^2/2$ is the energy density using the ideal equation of state: $\rho \epsilon = p/(\gamma-1)$ ($\gamma$ being the adiabatic exponent, and $\epsilon$ the internal energy per mass), $\mathbf{B}$ is the magnetic field, and $\mathbf{a}$ is a source term (gravitational acceleration in our case).

Our domain is $[R,Z] \in [0,100 R_*]\times[-100 R_*,100 R_*]$ with $768\times 1536$ grid points. We use a mixed grid (uniform+stretched) so that $256 \times 512$ grid points uniformly mesh the domain $[0,2.5 R_*]\times[-2.5 R_*,2.5 R_*]$, which surrounds the star. The grid spacing then grows geometrically with the distance to the star. 

The stellar wind solutions are obtained by setting boundary conditions at the border and inside the computational domain. Those boundary conditions are described  in the appendix \ref{AppB} and summed-up in Figure \ref{BC}.
PLUTO solves normalized equations. Three normalization values set all the others: length, speed and density. Thus in our set-up the radius of the star is $R_*/R_0=1$, the density and the keplerian speed at the surface of the star are $\rho_*/\rho_0=1$ and  $v_{kep}/V_0=\sqrt{GM_*/R_*}/V_0=1$. By choosing the physical values of those normalizations, for example $R_0=6.96 \times 10^{10}$ cm (the radius of the Sun) one can deduce all the other values output by the code. The magnetic field normalization is for instance given by : $B_0=\sqrt{4 \pi \rho_0}V_0$. Our simulations are then controlled with 5 parameters: $\gamma$ the adiabatic exponent (ratio of specific heats), the initial magnetic field topology which can be a dipole, a quadrupole or an octupole, and three speeds normalized by the escape speed, taken at the equator and at the surface of the star: $v_{rot}/v_{esc}$ the rotation speed, $v_{A}/v_{esc}$ the surface Alfv\'en speed at the equator (directly related to the strength of the magnetic field), $c_s/v_{esc}$ the speed of sound (giving the pressure over density ratio since $c_s=\sqrt{\gamma p /\rho}$).

We then let the code evolve the set of equations (\ref{MHD_1}-\ref{MHD_4}) until it reaches a steady-state solution. We check the quality of this steady state with various criteria. For instance, by looking at the mass flux versus time, one can be sure that a constant value is reached. Another method first introduced in \citet{KG1999}, is to look at several quantities that should be conserved along the field lines in ideal MHD. We used this technique with several quantities, especially the effective rotation rate. More details are given in the appendix \ref{AppB}.

Our purpose here is to investigate the effect of the magnetic topology on the magnetic braking of sun-like stars. We use for this the same method as the one developed in \citet{MB2004,MP2008,Matt2012}, \textit{i.e.} 2.5D axisymmetric ideal MHD simulations. However a different code is used to compute the wind solutions: PLUTO \citep{Mignone2007}. We kept fixed $\gamma$ and $c_s/v_{esc}$, at the fiducial values for sun-like stars \citep{Matt2012,WashShib1993}. The parameter $c_s/v_{esc}=0.222$ corresponds to a $\sim 10^6 $ K hot corona for solar parameters and $\gamma=1.05$. This choice of $\gamma$ is dictated by the need to maintain an almost constant temperature as the wind expands, which is observed in the solar wind. Hence choosing $\gamma \neq 5/3$ is a simplified way of taking into account heating which is not modeled here. For combined values of $v_A/v_{esc}$ and $f$ we chose 20 cases from \citet{Matt2012}. For each of these cases we run three different magnetic topologies for the star. The value of the parameters are summed-up in Table \ref{inpartable}. For a solar mass and radius, the range of value for the rotation period goes from 1167 days till 0.3 days approximately (from $10^{-4}$ to $0.4$ in terms of break-up ratio), while the strength of the magnetic field at the equator (controlled by $v_A/v_{esc}$) goes from 0.9 to 35 Gauss for a base coronal density of $2.9 \times 10^{-15}$ g/cm$^3$. However, changing normalizations naturally changes the physical parameter range (see section \ref{discuss}).

\begin{deluxetable}{lllccc}
  \tablecaption{Table of parameters and computed Alfv\'en radii\label{inpartable}}
  \tablecolumns{6}
  \tabletypesize{\scriptsize}
  \tablehead{
    \colhead{Case}&
    \colhead{$v_{A}/v_{esc}$} &
    \colhead{$f$} & 
    \colhead{$\langle R_A \rangle$ Dip.} & 
    \colhead{$\langle R_A \rangle$ Quad.} & 
    \colhead{$\langle R_A \rangle$ Oct.}
  }
  \startdata
  1&0.0753 &9.95e-5	&6.3	&3.6		& 3.0\\
  2&0.301	&9.95e-5	&12.5	&5.3	&4.0\\
  3&1.51	&9.95e-5	&32.3	&9.3		&5.9\\
  3+&2.00	&9.95e-5	&36.4	&9.9		&6.3\\
  5&0.0753 &9.95e-4	&6.3	&3.6		& 3.0\\
  6&0.301	&9.95e-4	&12.5	&5.3	&4.0\\
  7&1.51	&9.95e-4	&32.3	&9.3		&5.9\\
  8 &0.0753	&3.93e-3	&6.3	&3.6		&3.0\\
  10 &0.301	&3.93e-3	&12.6	&5.3	&4.0\\
  13 &1.51	&3.93e-3	&32.3	&9.3		&5.9\\
  23&0.0753	&4.03e-2	&5.9	&3.5	&3.1\\
  24&0.301	&4.03e-2	&11.7	&5.2	&4.1\\
  25&1.51	&4.03e-2	&30.3	&9.2	&5.8\\
  31&0.301	&5.94e-2	&10.6	&4.9	&4.0\\
  37&0.301	&9.86e-2	&8.7	&4.4	&3.6\\
  45&0.301	&1.97e-1	&5.5	&3.4	&2.9\\
  47&1.51	&1.97e-1	&13.4	&5.7	&4.4\\
  48&0.753	&4.03e-1	&4.9	&3.1	&2.6\\
  49&1.51	&4.03e-1	&7.1	&3.7	&3.2\\
  50&3.01	&4.03e-1	&11.4	&5.0		&4.3\\
  \enddata
  \tablecomments{Parameters of our 60 simulations (20 for each topology) with $\gamma=1.05$ and $c_s/v_{esc}=0.222$ in columns 2 and 3. The number the cases refer to \citet{Matt2012}. The resulting average Alfv\'en radii are given in the last 3 columns. They decrease with higher order topology, and with rotation starting at $f=0.04$ (cases 23-24-25).}
\end{deluxetable}

\begin{figure}
\center
\includegraphics[scale=0.7]{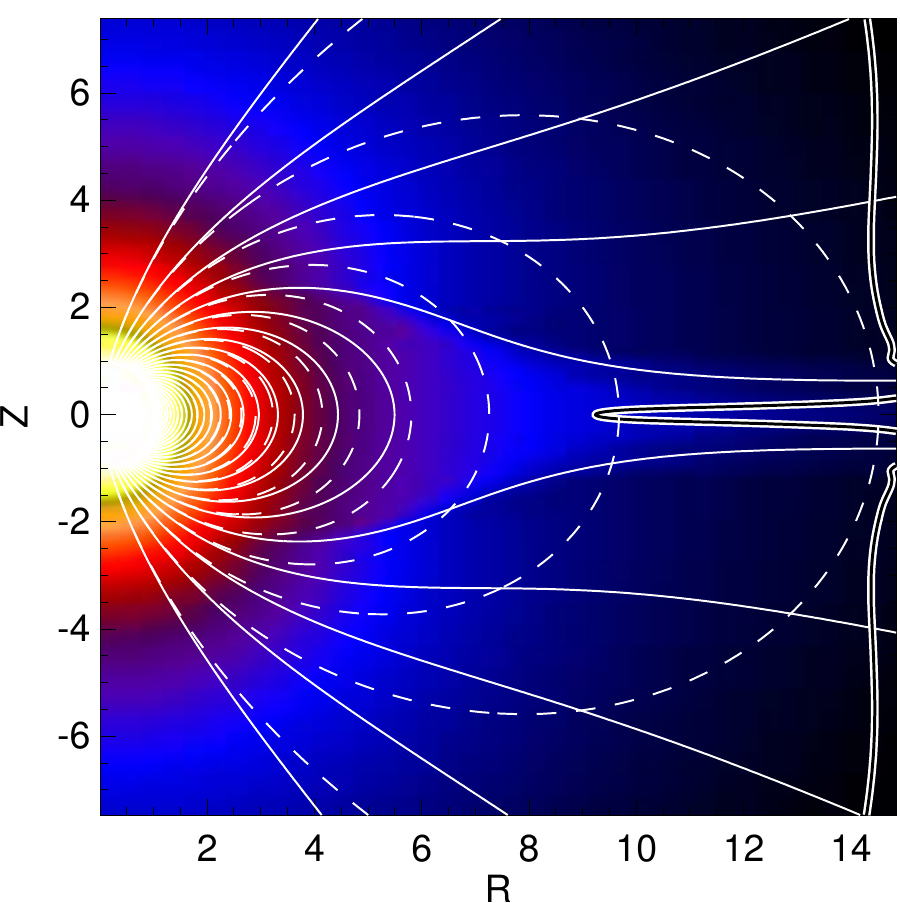} \\
\includegraphics[scale=0.7]{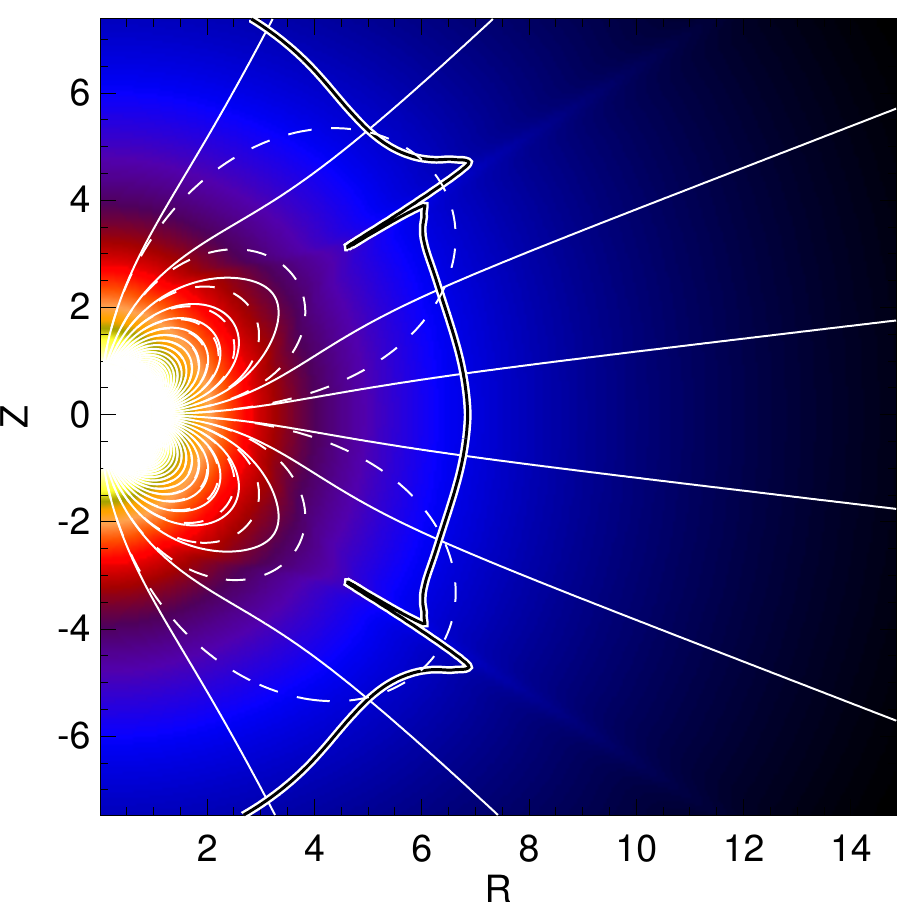} \\
\includegraphics[scale=0.7]{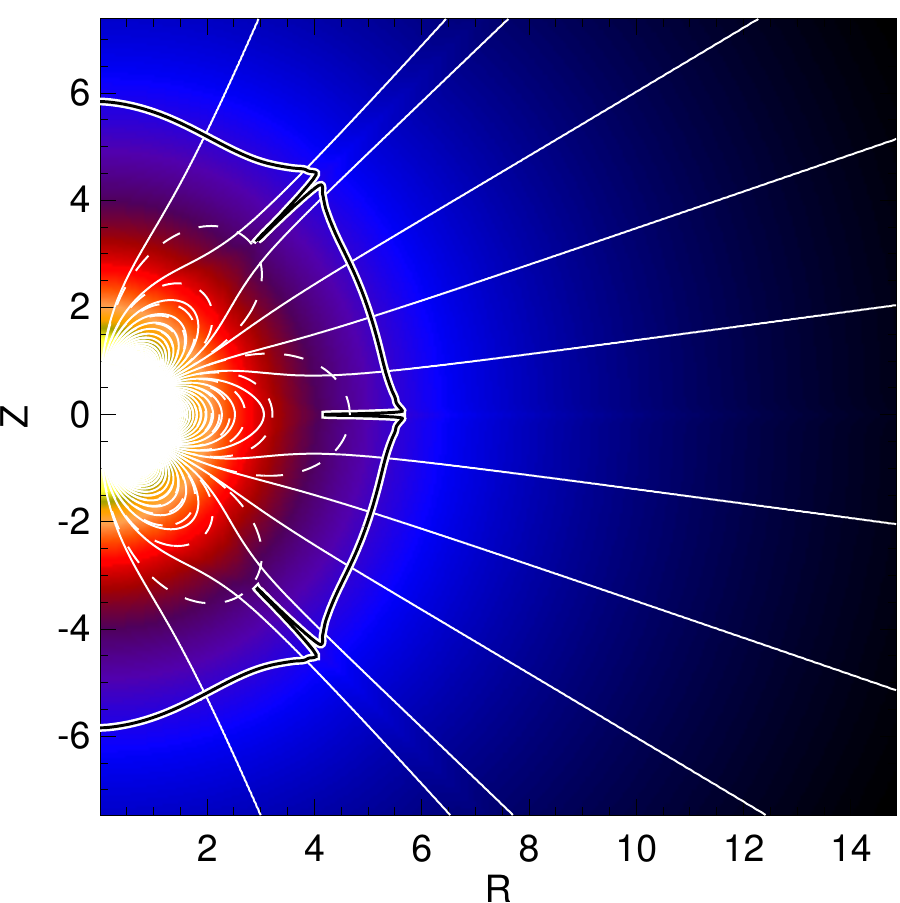}\\
\includegraphics[scale=0.24]{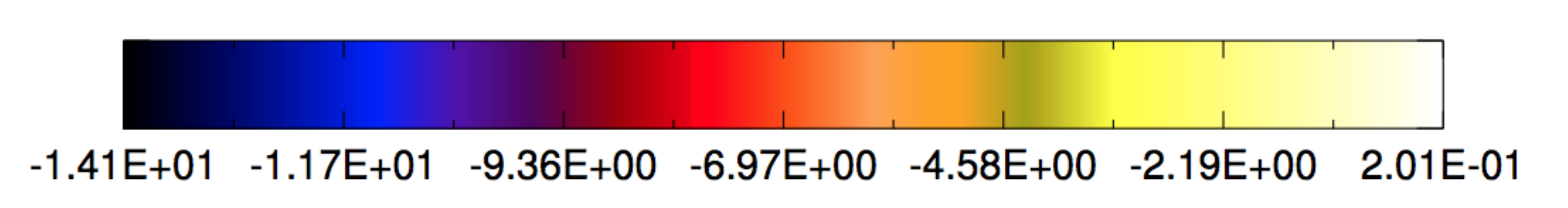}\\
\caption{Magnetic field lines of initial topologies (dashed lines) and final state (continuous lines) for case 2 and the three topologies. The color background is the logarithm of the density normalized to the surface density. The Alfv\'en surface is shown as the thick black and white line.}
\label{InitTopo}
\end{figure}

Figure \ref{InitTopo} shows these three topologies in the initial state (dashed field lines) and once a steady-state has been reached (continuous field lines). We now discuss the results of our 60 simulations.

\section{Parametric dependence of the magnetic torque}
\label{Parstudy}

\begin{figure*}[h!]
\center
\begin{tabular}{lll}
\includegraphics[scale=0.7]{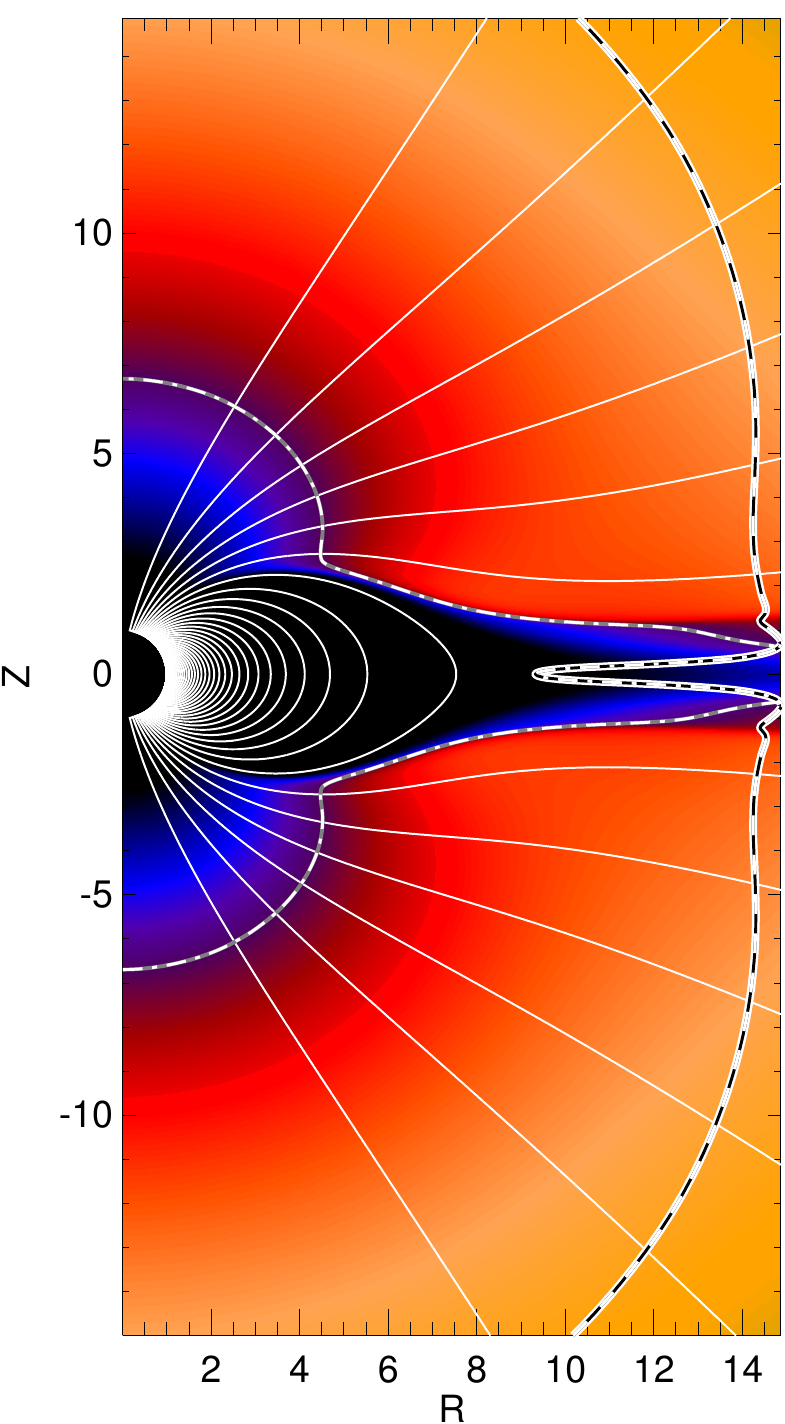}  & \includegraphics[scale=0.7]{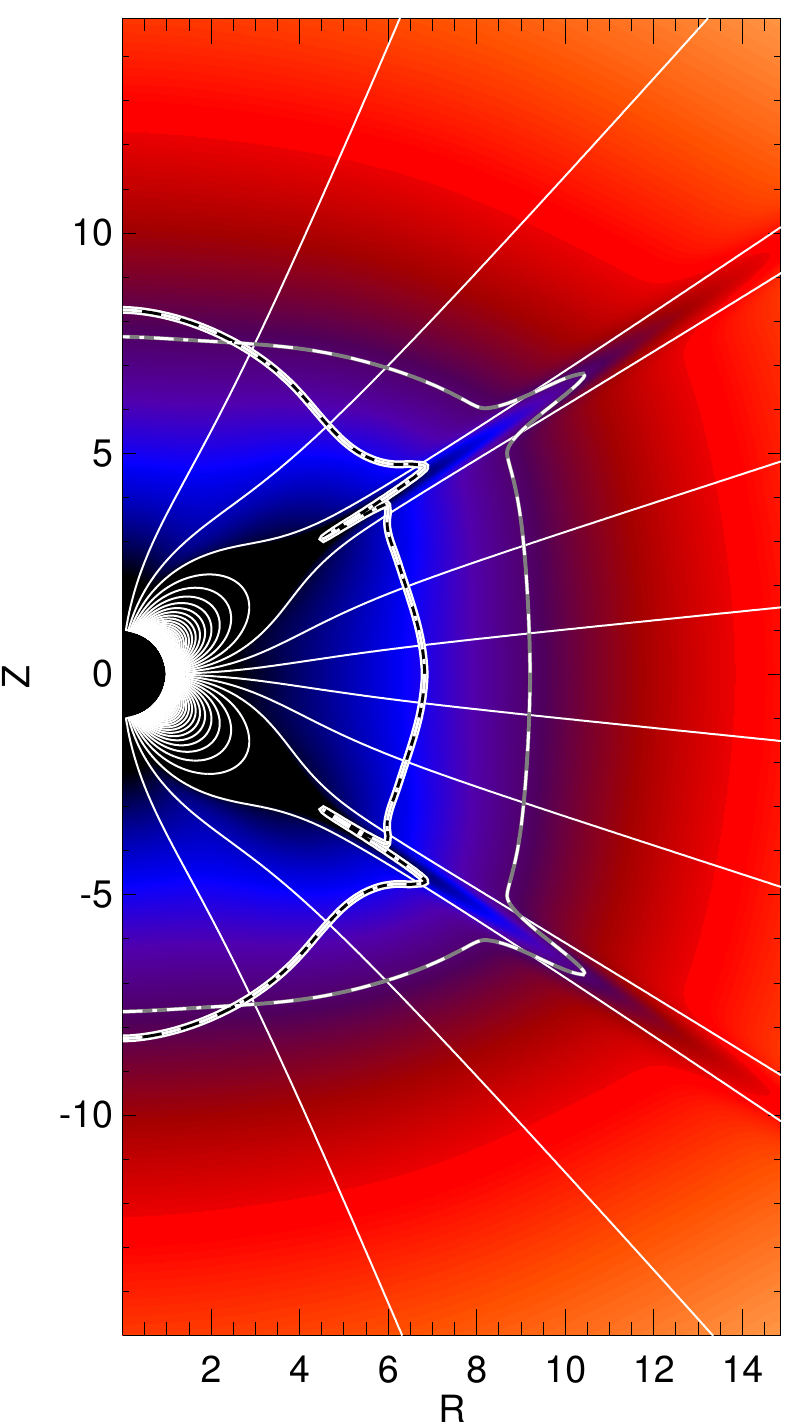} & \includegraphics[scale=0.75]{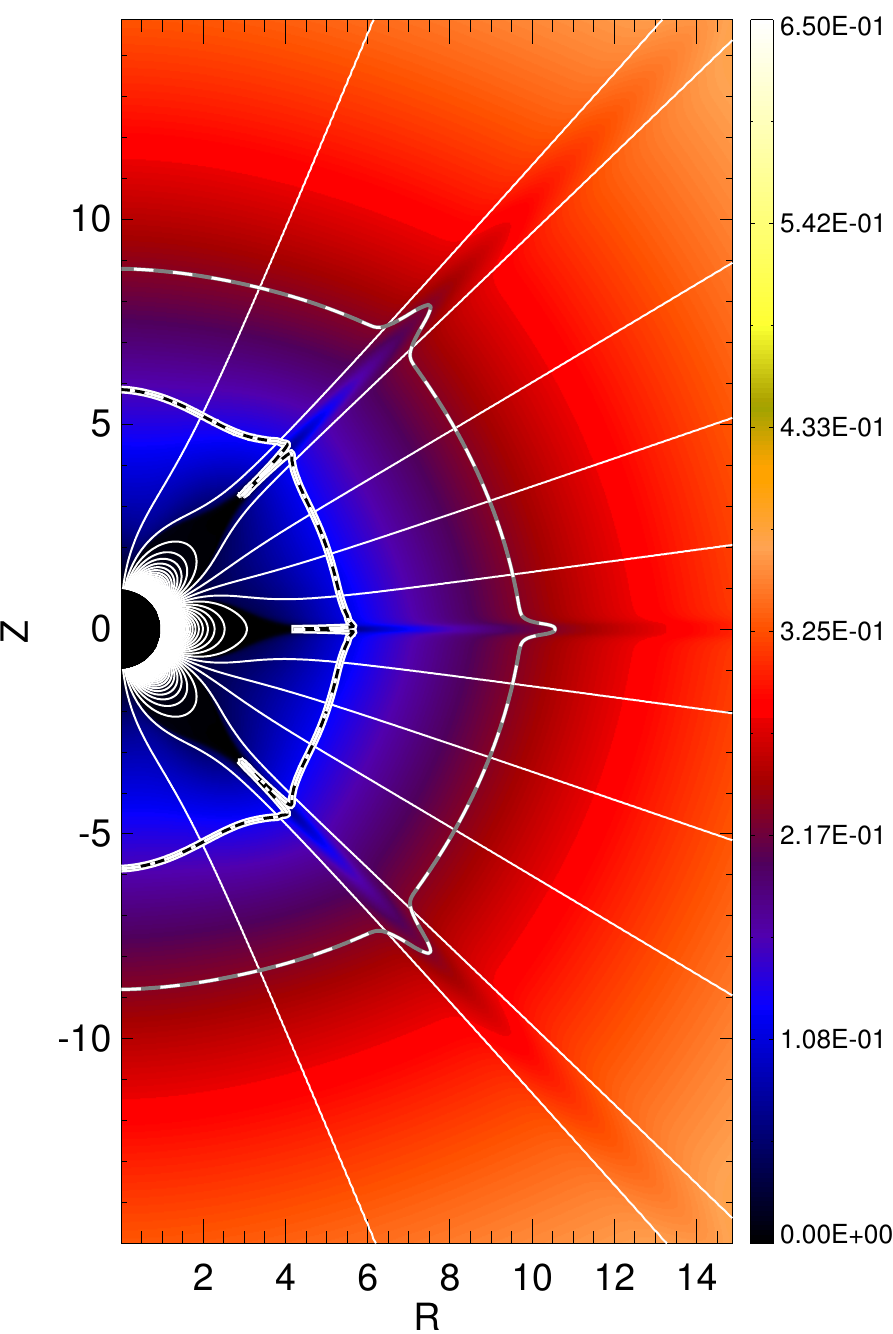} \\
\includegraphics[scale=0.7]{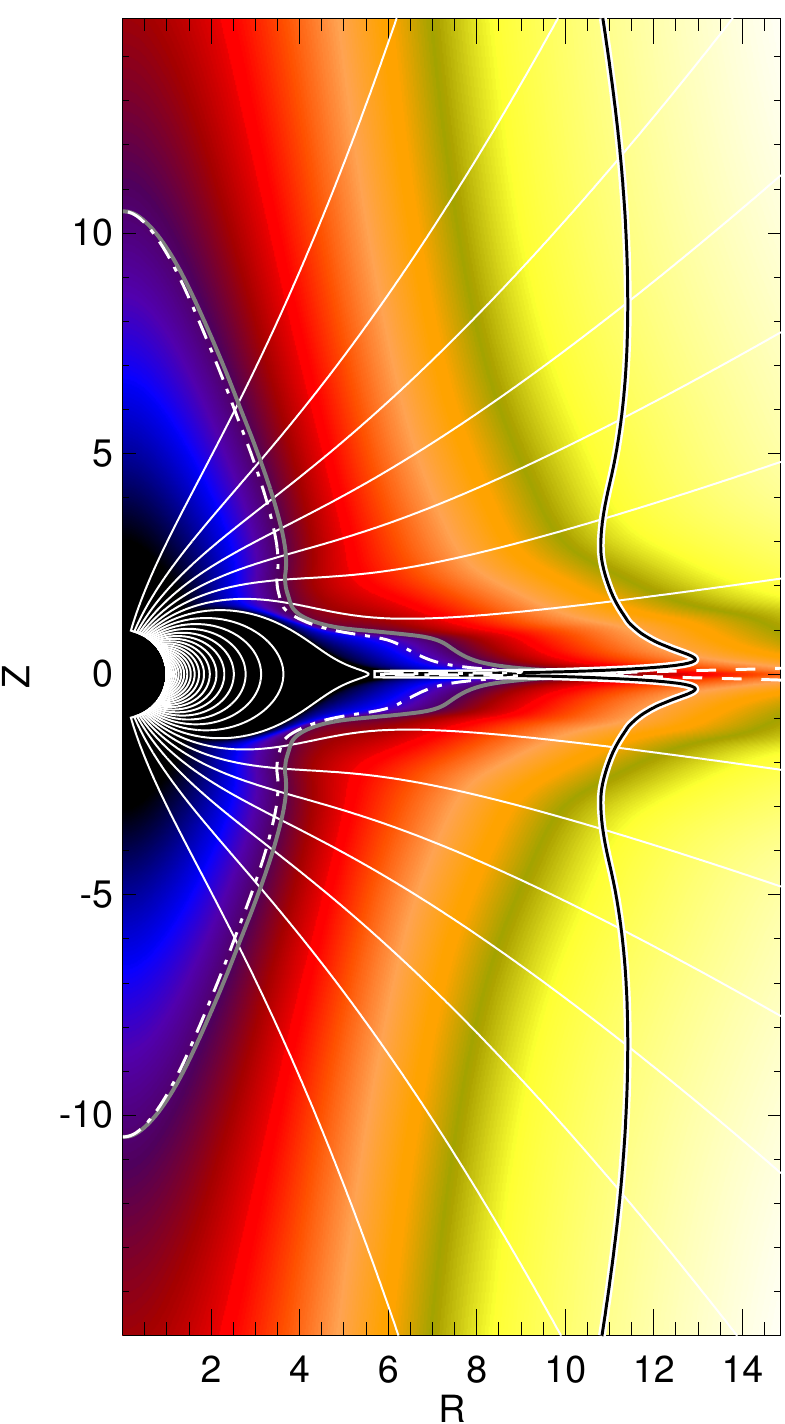} & \includegraphics[scale=0.7]{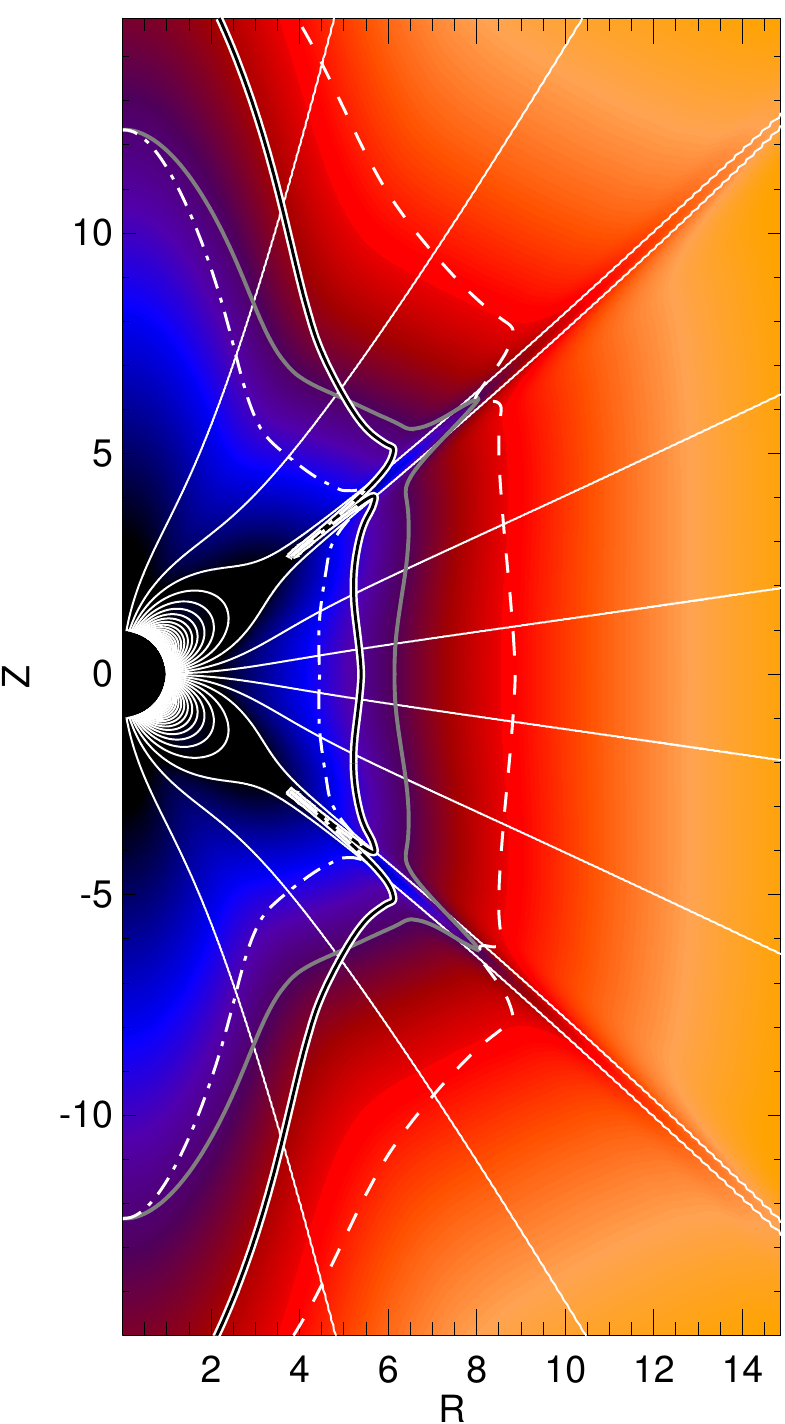} &  \includegraphics[scale=0.75]{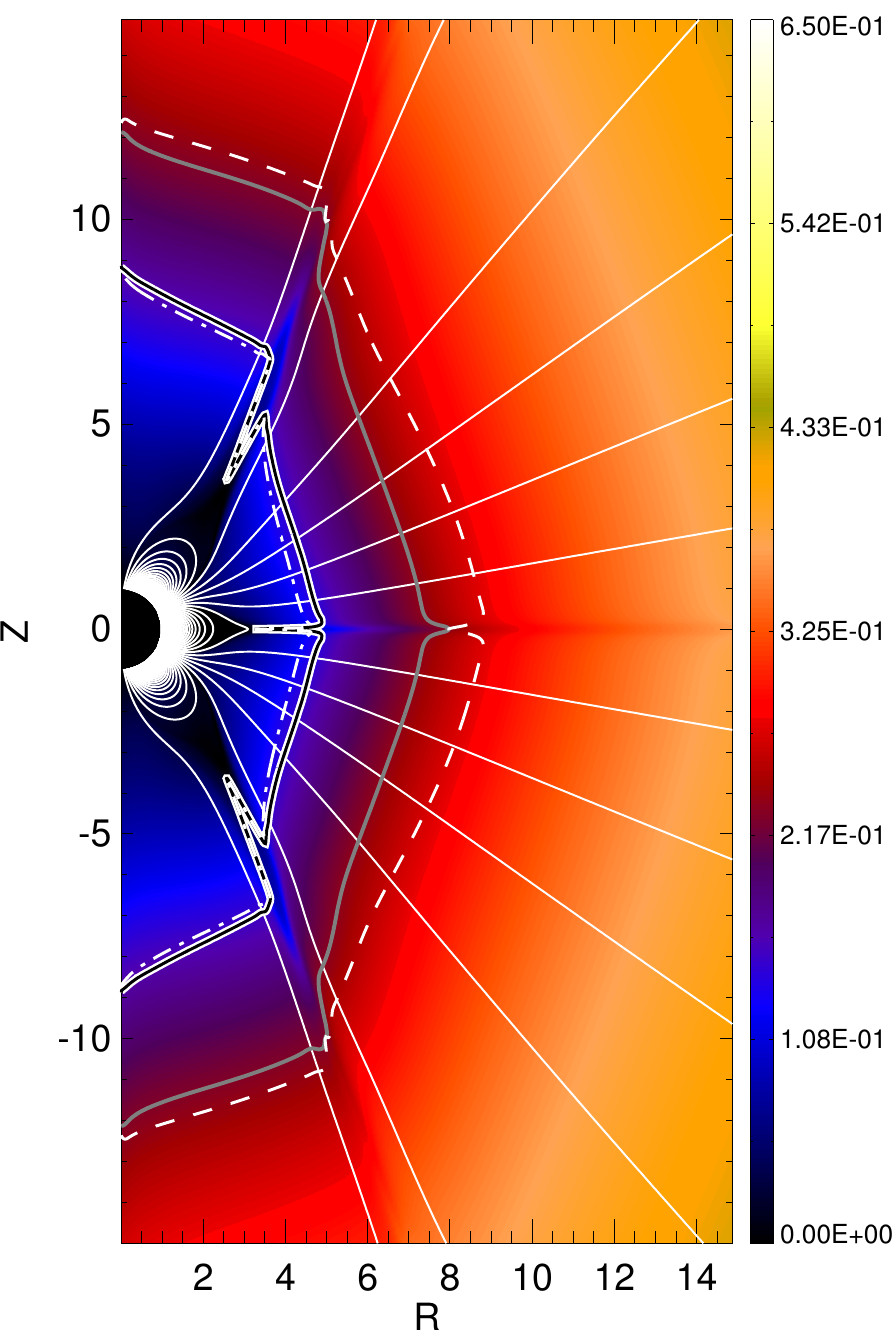}\\
\end{tabular}
\caption{Steady-state solutions for a wind with a dipolar, quadrupolar and octupolar magnetic field from left to right and for two different rotation rates: case 2 ($f=9.95\times 10^{-5}$) on top and 31 ($f=5.94 \times 10^{-2}$) below. The simulations are initialized with the same coronal temperature and the same magnetic field strength. The thick white line with a black core is the Alfv\'en surface and the thick grey line is the sonic surface. The slow and the fast magneto-sonic surfaces respectively are the dot-dashed and dashed lines. When rotation increases the Alfv\' en surface gets closer to the star at the equator and further at the pole. For higher order multipoles the Alfv\'en radius is also generally closer to the star. The poloidal speed normalized to the keplerian speed (437 km/s for the Sun) is represented by the color background with the same scale on all the panels.}
\label{WindVis}
\end{figure*}

Magnetic field topology has a strong influence on the outflow. Figure \ref{InitTopo} illustrates how the thermal and dynamical pressure of the wind opens the magnetic field lines. In steady state, magnetic loops remain centered on latitudes where the (spherical) radial magnetic field is zero, thus forming dead zones where the plasma co-rotates with the star. The shape and size of those dead zones depends on the magnetic field strength and the rotation rate, while their number is controlled by the magnetic topology. Generally speaking, we can extract trends for each independent parameter variation.

\subsection{Effect of magnetic field strength}
\label{sub:mag}
The magnetic field strength has quite a straightforward influence on the average Alfv\'en radius: a more intense magnetic field results in a larger torque (see Table \ref{inpartable}). This is because the Alfv\'en speed is higher, and thus it takes longer to the wind to accelerate and reach the Alfv\'en surface. The magnetic field strength also has a weak influence on the mass loss rate (compared to the rotation rate, see section \ref{discuss}). The latter decrease with stronger magnetic fields since magnetic forces are able to confine more plasma in the dead zones. However the Alfv\'en radius does not grow linearly with the magnetic field strength. For instance between case 2 and 3, magnetic field strength is multiplied by 5, whereas the average Alfv\'en radius is multiplied by 2.6 (see Table \ref{inpartable}).

\citet{MP2008} precisely described this effect, and for weakly rotating cases, we are able to fit the average Alfv\'en radius with the same formulation (equation \ref{simple_power_law}). More details on the fit will be given in section \ref{sub:rot}.

\subsection{Magneto-centrifugal regime and force budget (effect of rotation)}
\label{sub:rot}

In Figure \ref{WindVis} several solutions illustrate the influence of rotation and topology on the wind. For the three different topologies, the rotation rate is increased while the magnetic field strength is held fixed (top panel is case 2, bottom panel is case 31). We plot  the surfaces corresponding to each modes of ideal MHD along with the sonic surface. Fast and slow magneto-sonic surface are often merged with the sonic and Alfv\'en surfaces, and switch when those two cross, so that the fast magneto-sonic surface is always further from the star than the slow one. The poloidal speed (in color background) is strongly affected by the change in rotation rate in the bottom panels.

\begin{figure}[!h]
\center
\includegraphics[scale=0.55]{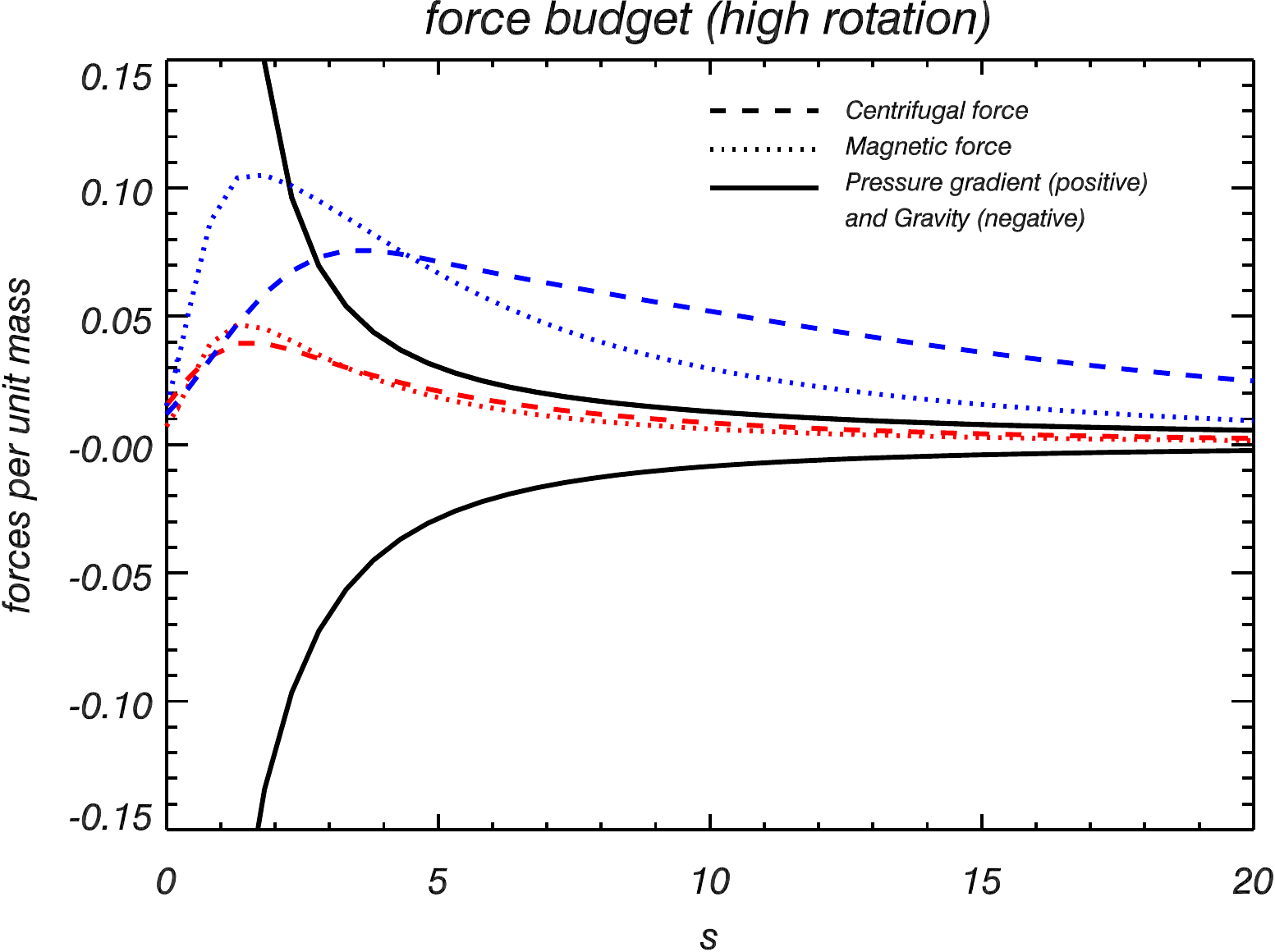}\\
\medskip
\includegraphics[scale=0.55]{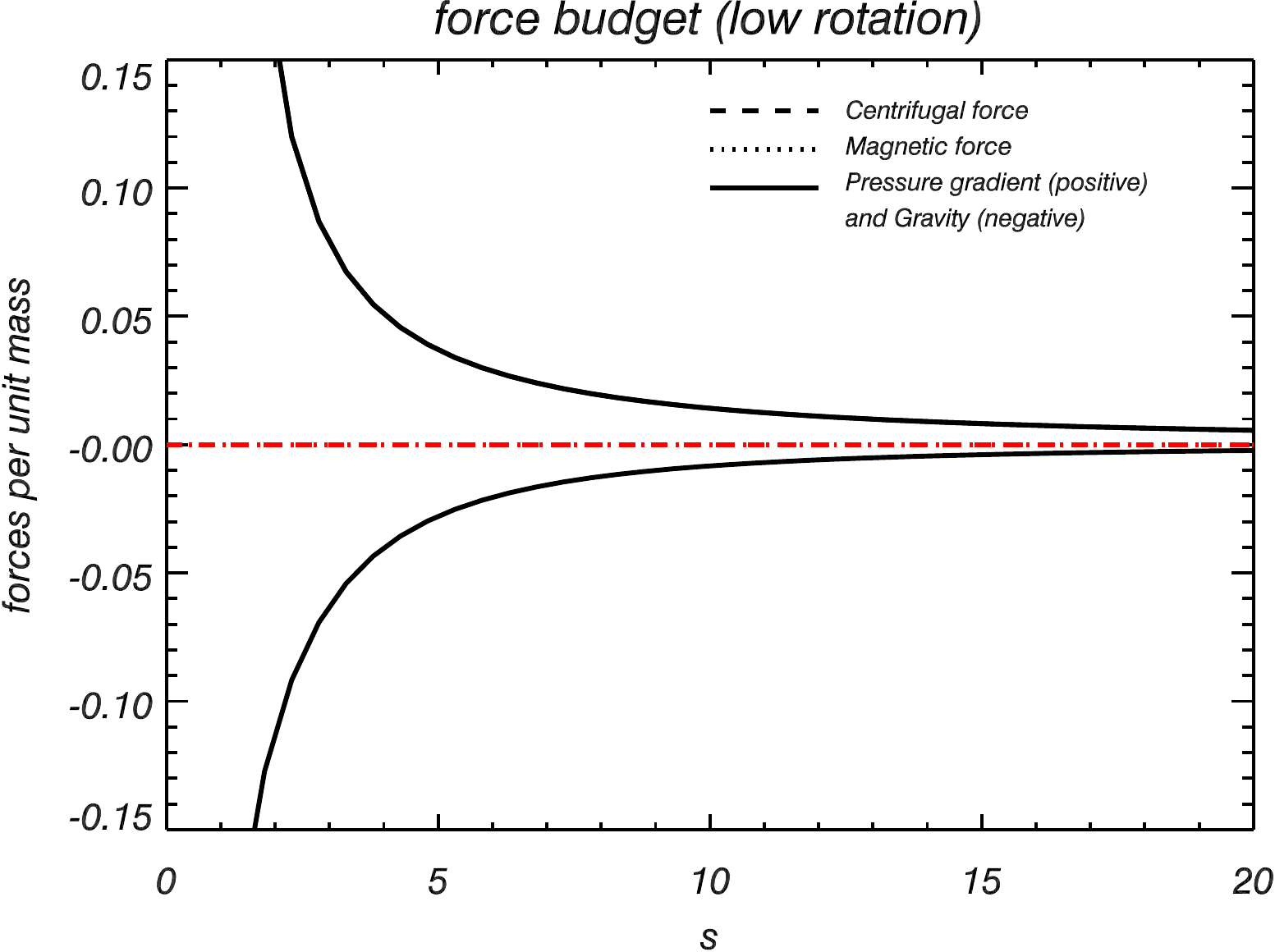}
\caption{On the two panels are the force budget along a field line for the dipolar cases 2, 45 and 47. The pressure gradient and gravity only remain close between the different cases, while the magnetic and centrifugal force go from negligible (top panel for case 2) to comparable (colored lines for case 45 (red) and 47 (blue) in the bottom panel) to the pressure gradient. The field lines have their seeds at 75$^{\circ}$, 55$^{\circ}$, 60$^{\circ}$ of latitude for cases 2, 45 and 47 respectively so that they remain approximately radial.  With higher magnetic field strength, the amplitude of the centrifugal and magnetic force increase. $s$ is the curvi-linear abscissa, expressed in stellar radii, that starts at the surface of the star.}
\label{forcebud}
\end{figure}

However the influence of the rotation rate appears through two different phenomena depending on latitude. First, and it is the most relevant aspect for angular momentum loss, the Alfv\'en surface comes closer to the star at low and mid latitudes wither higher rotation rate (see Figure \ref{WindVis} and Table \ref{inpartable}). This is a simple consequence of the magneto-centrifugal effect described by \citet{WeberDavis1967,Sakurai1985,WashShib1993}. The magnetized wind is rotating with the star and is thus accelerated by a centrifugal effect. A magnetic pressure gradient is also responsible for an additional acceleration. All forces projected along a magnetic field line can be expressed as follow \citep{Ustyugova1999}:

\begin{equation}
f_p=-\frac{1}{\rho}\frac{\partial p}{\partial s}
\end{equation}

\begin{equation}
f_g=-\frac{\partial \Phi}{\partial s}
\end{equation}

\begin{equation}
f_m=-\frac{1}{8\pi\rho R^2}\frac{\partial (R B_{\phi})^2}{\partial s}
\label{fm}
\end{equation}

\begin{equation}
f_c=-\frac{v_{\phi}^2}{R} \quad \mathbf{\hat{s}} \cdot \mathbf{\hat{R}}
\end{equation}

where the subscripts $p$, $g$, $m$, $c$ refer respectively to the pressure gradient and the gravitational, magnetic and centrifugal forces, while $s$ is the curvi-linear abscissa directed along the field line and $\Phi$ the gravitational potential. Vectors with a hat are unit vector. The force budget in slow and fast rotators is given in Figure \ref{forcebud}. One can see that at high rotation (20 \% of the break-up speed) the magnetic and centrifugal forces become comparable to or even higher than the pressure gradient. There is creation of a toroidal component of the magnetic field which contributes to the driving (equation \ref{fm}). This additional acceleration can double the poloidal speed around the equatorial streamer (for the highest values of $f$, see Figure \ref{WindVis}) and thus allow the outflow to reach the Alfv\'en speed much closer to the star. The transition between those two regimes (thermal and magneto-centrifugal winds) occurs around $f \approx 10^{-2}$ when magnetic and centrifugal forces reach a few percents of the pressure gradient.

Second, rotation extends the Alfv\'en surface at high latitudes. Rotation twist the magnetic field lines, through creation  of a toroidal component of the magnetic field. Near the poles the associated $B_{\phi}$ gradient is directed towards the rotation axis \citep{Ustyugova1999}. The magnetic field lines, being frozen-in the plasma, are bent as well and this collimation increases the poloidal field relative to the non-collimated case. Indeed rotation tend to decrease the width of a given flux tube near the poles and thus flux conservation imposes a poloidal field increase. The Alfv\'en surface then goes further away at high latitudes while poloidal speed is much less influenced by the centrifugal acceleration. In our simulations the Alfv\'en surface can be pushed further away by 250\% near the axis. For instance in dipolar case 31 the Alfv\'en surface is reached around 70 $R_*$ on the axis whereas in dipolar case 2 it is reached around 20 $R_*$ (the window of Figure \ref{WindVis} is too small to see those effects). However this effect does not seem to have a strong influence on the average global value $\langle R_A \rangle$ since angular momentum flux is relatively weak near the rotation axis (see Table \ref{inpartable}).

\begin{figure}[h!]
\center
\includegraphics[scale=0.55]{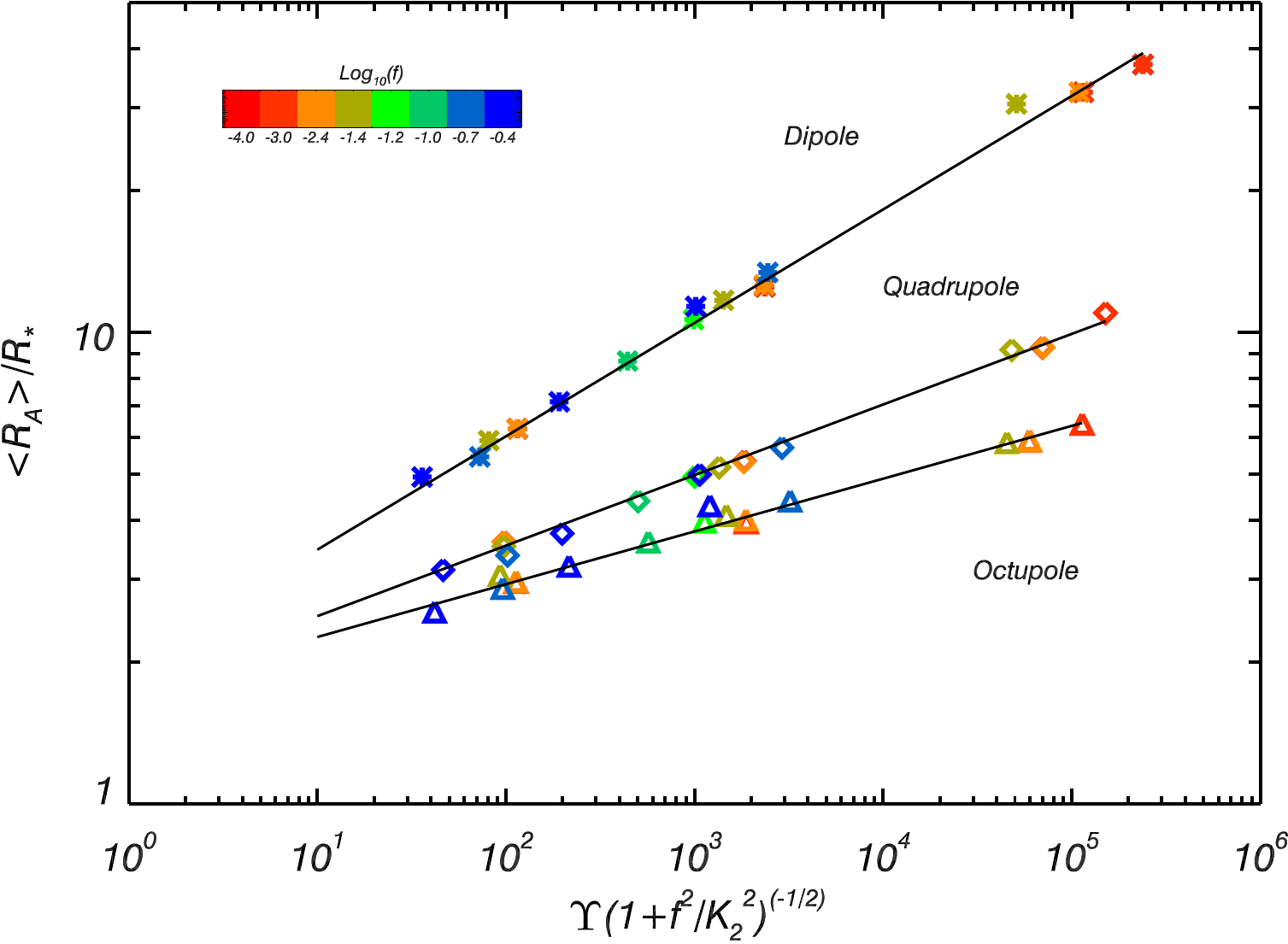}
\caption{The three fits made for each topology (parameters are in Table \ref{threetoppar}). The crosses are for the dipole, the diamonds for the quadrupole and the triangles for the octupole. Colors go from red to blue when rotating faster. With growing complexity of the magnetic field, the braking efficiency decreases as shown by the values of the fits.}
\label{threetop}
\end{figure}

In our parameter study (see Table \ref{inpartable}), the effect of rotation on the average Alfv\'en radius starts to be noticeable with cases 23-24-25 with a break-up ratio $f=0.04$ for all topologies. There magnetic and centrifugal forces reach a few percents of the pressure gradient. Up to $f \approx 0.1$, this effect is still weak and  the simple power-law (equation \ref{simple_power_law}) provide a good fit. Beyond this value ($f > 0.1$) the decrease of the Alfv\'en radius becomes stronger, justifying a three parameter regression, as in \citet{Matt2012}. Formulation \ref{form1} describe accurately our results and Figure \ref{threetop} illustrates this fit made independently for the three topologies (three black lines). All the results and outputs of our simulations are given in Table \ref{restable} while the results of the fits are in Table \ref{threetoppar}. 

\begin{deluxetable*}{lccccc}
\tablecaption{Fit parameters for the three topologies.\label{threetoppar}}
  \tablecolumns{6}
  \tabletypesize{\scriptsize}
  \tablehead{
    \colhead{Topology} &
    \colhead{$K_1$} & 
    \colhead{$K_2$} & 
    \colhead{$m$} &
    \colhead{$m_{th}(q=0.7)$}&
    \colhead{$m_{th}(q=-1/2)$}
  }
  \startdata
  Dipole ($l=1$) & $2.0 \pm 0.1$ & $0.2 \pm 0.1$ & $0.235 \pm 0.007$ & 0.21 & 0.29 \\
  Quadrupole ($l=2$) & $1.7 \pm 0.3$ & $0.2 \pm 0.1$ & $0.15 \pm 0.02$ & 0.15 & 0.18\\
  Octupole ($l=3$) & $1.7 \pm 0.3$ & $0.2 \pm 0.1$ & $0.11\pm 0.02$  & 0.11 & 0.13\\
  Radial/Open Field ($l=0$) & & & & 0.37 & 0.66 \\
   \hline
   \vspace{-0.1cm} & & & & \\ 
  
  & $K_3$ &$K_4$& $m$ & - & - \\
  \hline 
  \vspace{-0.1cm} & & & & \\ 
  Topology Independent & $1.4 \pm 0.1$ & $0.06 \pm 0.01$ & $0.31 \pm 0.02$  && \\
   \enddata
  \tablecomments{Parameters of the fit to equation \ref{form1} made independently for each topology. The values $K_1$, and $m$ corresponds to three black lines of Figure \ref{threetop} for the dipolar, quadrupolar and octupolar configurations. The parameter $K_3$ and $K_4$ for the topology independent formulation \ref{form2} (see Section \ref{Collapse}) are given as well. Expected value for $m_{th}$ through analytical models are given for different values of the parameter $q$ for comparison with the fitted value (see Section \ref{parvalues} and Appendix \ref{AppA}).}
\end{deluxetable*}

\subsection{Effect of topology}
\label{sub:top}

Qualitatively, in all three topologies we observe streamers at latitudes where magnetic forces are high enough to counterbalance the thermal and the ram pressure. Around the dead zones the plasma is moving slower than in open field lines regions (see Figure \ref{WindVis}). The magnetic field reaches a minimum (in ideal MHD it should be zero, and the Alfv\'en surface should touch the last closed magnetic field loop) on top of the dead zones while on both sides of this minimum the Alfv\'en surface is slightly extended due to the slow wind. At those latitudes, beyond dead zones, thin current sheets are created, one for the dipole, two for the quadrupole and three for the octupole. 

Interestingly the value of $\Upsilon$ is usually comparable (same order of magnitude) for the three topologies. Colored points of Figure \ref{threetop} are almost vertically aligned, which means that the mass loss rate does depend weakly on the magnetic topology (see equation \ref{Ups_def}). However a trend for the effect of topology on the mass loss can be detected with the strongest magnetic fields (see Table \ref{restable}), where lower order topologies have higher $\Upsilon$ (see Appendix \ref{AppC}).

\begin{figure}[!h]
\center
\includegraphics[scale=0.55]{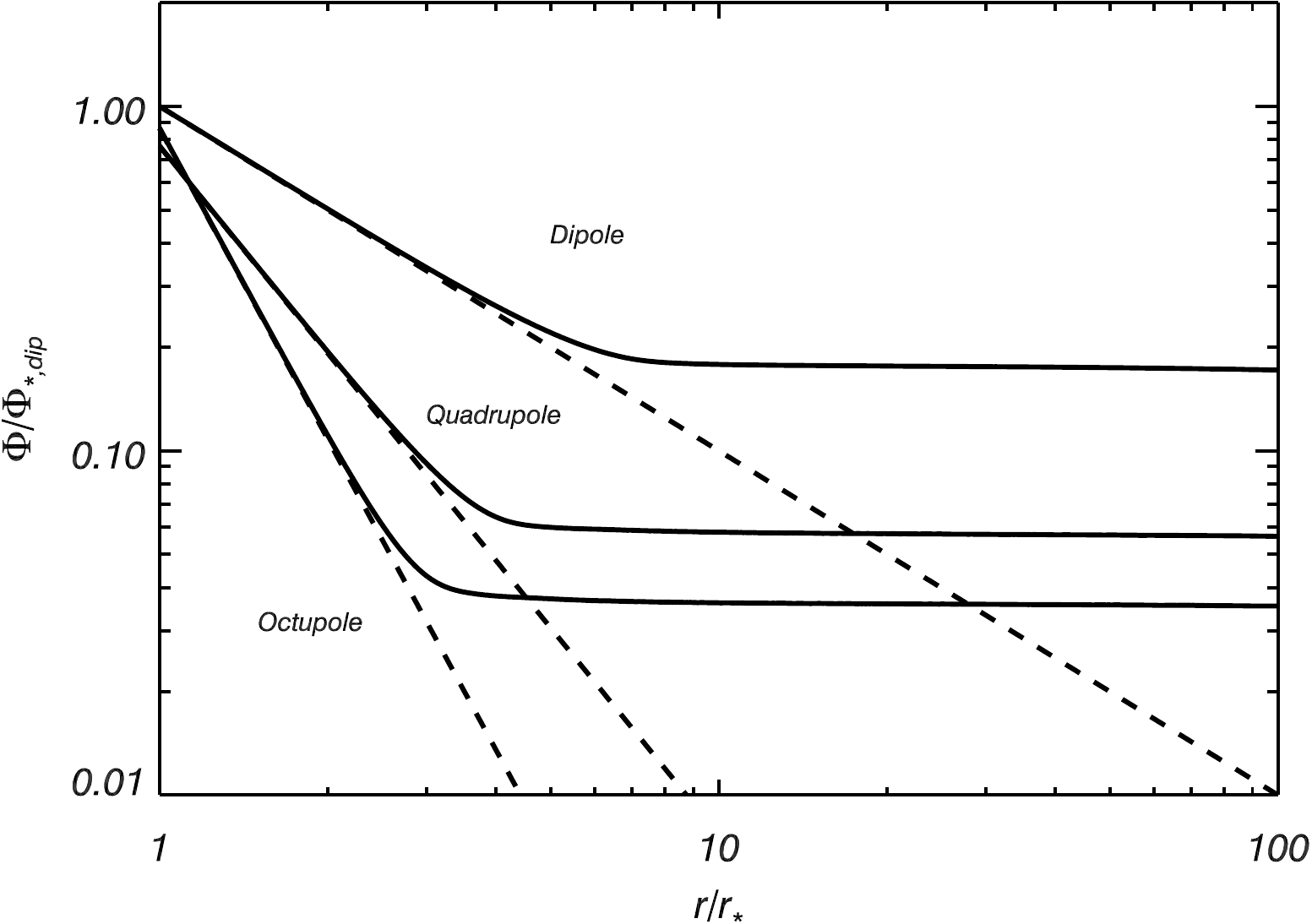}
\caption{Magnetic flux for all three topologies in case 3. The black lines are the magnetic flux in the steady-state winds while the dashed lines are the fluxes of the initial potential field as a function of the distance from the star.}
\label{threeflux}
\end{figure}

The key point of our parameter study is that with an increasing complexity of the topology (higher order multipole), the magnetic braking is less efficient for a given Alfv\'en speed at the base of the corona (see Figure \ref{WindVis} for relative positions of the Alfv\'en surfaces and Table \ref{inpartable} for the average value). Dipolar cases are always above quadrupolar cases, which are always above octupolar cases, in Figure \ref{threetop} for a given $\Upsilon$. The slope of the fits also decreases with higher order of multipoles (see Table \ref{threetoppar}). This is a consequence of the faster decay with distance to the star of the magnetic field for more complex topologies. This decay is shown in Figure \ref{threeflux}, where the absolute magnitude of the magnetic flux is integrated over concentric spheres.

\begin{equation}
\Phi(r) = \int_{S_r} |\vec{B} \cdot d\vec{S}|
\end{equation}

As shown in Figure \ref{threeflux}, The magnetic flux shows two regimes. Close to the star the magnetic flux decreases as $1/r^l$ where $l$ is 1 for the dipole, 2 for the quadruple and 3 for the octupole. In other words the flux follows radial dependence expected for single-mode topology. However, the wind opens the field lines and beyond some stellar radii, the field becomes completely open, and thus the flux becomes constant. This constant is the value we define as the open flux, which is numerically computed from our simulations (see section \ref{Collapse}). The decay of the three modes is such that even if the field lines open closer to the star in higher order multipole, the amount of magnetic flux is higher for lower order multipoles. Hence the Alfv\'en speed is logically reached closer for more complex topologies since the amplitude of the magnetic field is still lower.

In our 60 simulations, we chose to use the parameter $v_A/v_{esc}$ to control magnetic field strength. This parameter is proportional to $B_*$ which is the magnetic field amplitude taken at the surface of the star, on the equator. Thus, for different topologies, the magnetic surface flux varies slightly at $r/r_*=1$ as it can be seen on Figure \ref{threeflux}. The variation is smaller than 29\% when comparing the various topologies used in our study, independently of the value of $B_*$\footnote{The flux at the surface of the star is $\phi(R_*)=\alpha \pi B_*R_*^2$ where $\alpha=4, 16/(3\sqrt{3})\approx 3.1 \text{ and }   52/15 \approx 3.5$ for the dipole, the quadrupole and the octupole respectively.}. Hence a parametric study that varies the surface magnetic flux would give similar braking laws.  On the other hand the value of the open flux is a complex consequence of the dynamics of the wind, and topology has a strong influence on it (see Figure \ref{threeflux}). We will show in the next section how the open flux can be used to derive a topology independent braking law.

\section{Towards a general braking law}
\label{Collapse}

\subsection{A topology-independent formulation for the magnetic torque}
\label{TopoInd}

Topologies of cool stars' magnetic fields include combination of dipolar, quadrupolar and octupolar components as well as higher order multipoles. This configuration changes over magnetic cycles that are likely to occur in most solar-like stars (see \citet{PintoBrun2011} for the effect of the 11 year cycle variability of the Sun on the wind topology). We attempt to find a topology-independent formulation in order to take into account this complexity for stellar evolution models. This section gives a single law fit for our 60 simulations and all three topologies. The key idea is to consider the dependency of the Alfv\'en radius on the open magnetic flux instead of the surface magnetic field strength. In all our cases the open flux is the constant value of the integrated magnetic flux beyond the last magnetic loops (see previous section, Figure \ref{threeflux}). \\

\begin{figure}[h!]
\center
\includegraphics[scale=0.55]{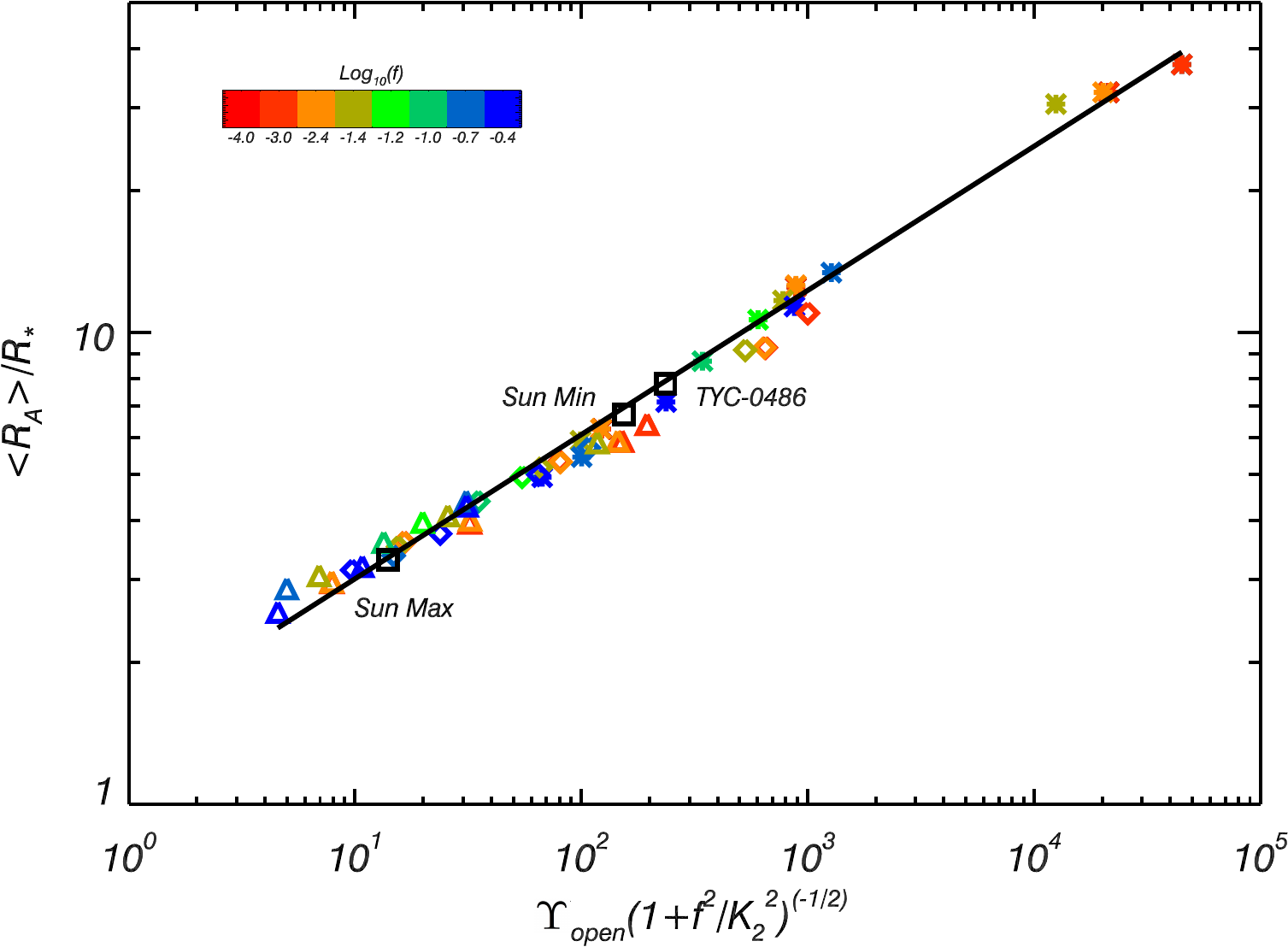}
\caption{The one law fit for the three topologies. Colors and symbols are the same as Figure \ref{threetop}. The influence of the rotation is captured in x-axis variable as in Figure \ref{threetop}. Black squares are complex topologies (Sun Min, Sun Max and TYC-0486-4943-1 introduced in section \ref{Realistic}), which demonstrates that winds with combination of magnetic multipoles follow the same braking law as those with single-mode topologies.}
\label{Col3par}
\end{figure}

Thus introducing the open flux into a new parameter $\Upsilon_{open}$ defined as follow:

\begin{equation}
\Upsilon_{open} \equiv \frac{\Phi_{open}^2}{R_*^2 \dot{M}_w v_{esc}},
\end{equation}

and using a similar formulation to the previous section:

\begin{equation}
\frac{R_A}{R_*}=K_3 [\frac{\Upsilon_{open}}{(1+f^2/K_4^2)^{1/2}}]^m,
\label{form2}
\end{equation}

we are able to fit all our 60 simulations into one single law, as shown in Figure \ref{Col3par}. The value for the parameters of this law are in Table \ref{threetoppar}.\\

Over the 60 simulations the average deviation to the fit is around 4\% while 6 cases reach between 15\% and 20\%. This formulation shows how relevant is the open flux in this situation. Indeed the mass and angular momentum losses occur through open field lines. And $\Upsilon_{open}$ contains the relevant information about the open field line region and the size and the number of dead zones, since these determine how much open flux there is, compared to the surface magnetic flux. As a consequence the open flux and the mass loss rate are coupled \citep[see][]{Vidotto2014}. Higher mass loss corresponds to higher open flux for a given $B_*$, that is to say rotation increases both. Thus the effect of rotation can be seen earlier in the open flux formulation (the fitted value of $K_4$ is $0.06$ rather than $K_2=0.2$ in section \ref{Parstudy}). We show in Figure \ref{Col2par} (in Appendix \ref{AppC}) how important is to introduce a three parameter fit, taking into account the rotation rate. Comparing Figure \ref{Col2par} and \ref{Col3par} we see that the denominator $(1+f^2/K_4^2)^{0.5}$ in formulation \ref{form2} collapses all the points on the single power-law.

\subsection{Comments}
\label{parvalues}

In our work we have been through 20 cases of \citet{Matt2012} parameters. Comparing both works, we can see that our values for $\Upsilon$ are in all cases larger, sometimes by an order of magnitude. Different computed mass loss rates are responsible for this mismatch. Our trend for higher $\Upsilon$ and thus lower mass loss rate can be understood. We implemented a somewhat different boundary condition that sets the pressure and density gradients inside the star using a polytropic (rather than isothermal) hydrodynamical wind solution \citep{KG2000}. This makes our simulation more stable numerically since an inherent irregularity of the speed solution at the surface of the star is removed thanks to this method. Indeed in \citet{Matt2012} the wind is isothermal inside the star and polytropic outside. And for $\gamma=1.05$ the speed profile is below the isothermal solution at a given radius. Even though the outside domain is ruled by a polytropic equation of state in both works, the isothermal boundary condition seems to maintain the steady state solution with a higher $\dot{M}_w$ than the fully polytropic simulation.

In the end, our parameter study gave us three topology associated braking laws, using the same formulation as \citet{Matt2012} but with different coefficients, and a topology independent formulation. For the dipolar cases, comparing our results to \citet{Matt2012}, it is interesting to note that despite very different values of $\Upsilon$, the exponents $m$ for the dipolar case, which are 0.235 in our case and 0.22 in \citet{Matt2012}, are almost statistically indistinguishable. This show that even if the prediction of $\dot{M}_w$ is highly sensitive to numerical and thermodynamical properties of the simulations, $\Upsilon$ is the relevant control parameter for stellar wind braking.

Interestingly, the value of $K_1$ and $K_2$ do not vary much with the topology. The value of $K_1$ is somewhat constrained by a similar behavior of the Alfv\'en surface when $\Upsilon \sim 1$, \textit{i.e.} for very weak magnetic fields. The coefficient $K_2$ is logically similar since the same acceleration process occur in the open field lines with the three topologies. However the topology has a strong influence on the braking through the $m$ coefficient and to take into account the complex magnetic field of solar-like stars in a single equation, the new formulation we propose is needed.

Our dipolar value for $K_1$ is a bit smaller than the one obtained in \citet{Matt2012}. We think that this value is very sensitive to our change of boundary conditions. It is the most likely to change with the thermal driving, \textit{i.e} the value of $\gamma$ and $c_s/v_{esc}$ that remained fixed in our parameter study (see \citet{MP2008} for some cases with different $c_s/v_{esc}$ and $\gamma$, and section \ref{Mdot}).

For the three topologies our value for the $K_2$ constant is larger ($\sim 0.2$ rather than $0.07$) from the one found in \citet{Matt2012}. However, as we see in Table \ref{inpartable}, rotation starts to influence the average Alfv\'en radius around $f = 0.04$, but the errors are significant in those parameters and we have less coverage of the rotation rate than in \citet{Matt2012} . The difference might also be due to the slight changes of boundary conditions in comparison with this work.

The parameter $K_4$ that appears in the open flux formulation seems to be the most relevant to consider the influence of the rotation rate. The value of $0.06$ is coherent with our forces analysis. It is also very close to the one obtained in \citet{Matt2012} ($K_2=0.0716$), and in \citet{WashShib1993} where it is also found that above $f=0.079$ the centrifugal effect begins to be prominent.

\begin{figure}[!h]
\includegraphics[scale=0.56]{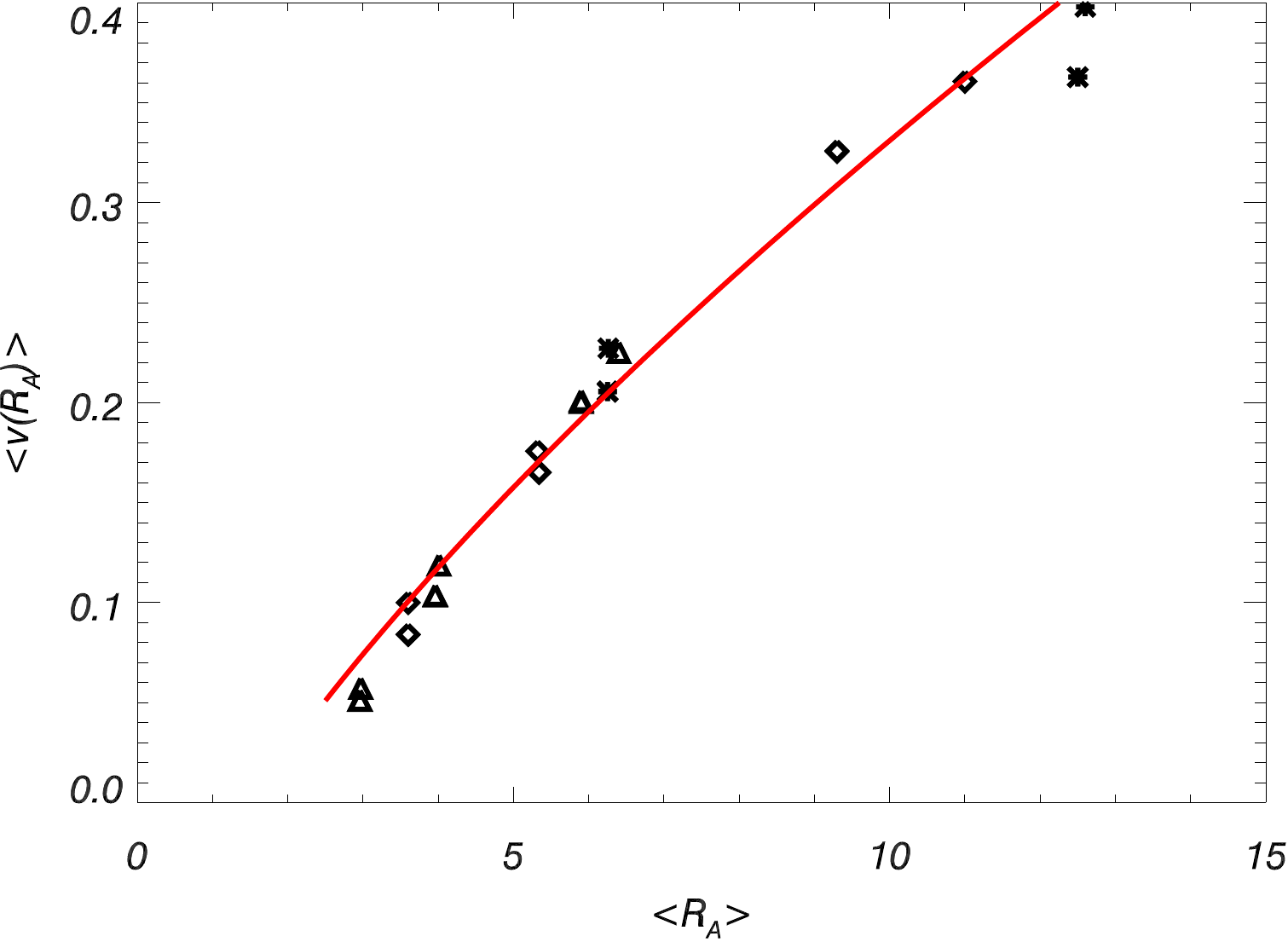}
\caption{Scatter plot of the average Alfv\'en speed on the Alfv\'en surface as a function of the average Alfv\'en radius (corresponding to equation \ref{vra}) for 26 slow rotating cases. The red line is a fit of a power law growth with an exponent $q=0.7$ and a slight offset.}
\label{AlfSpeed}
\end{figure}

For the exponent $m$ of the power law, analytical calculations give $m_{th}=1/(2l+2+q)$, where $1/r^{l+2}$ is the radial dependency of the magnetic field and $q$ is the exponent of the power law that describe the Alfv\'en velocity dependency on the Alfv\'en radius (see Appendix \ref{AppA}).
\citet{Kawaler1988} and \citet{ReinersMohanty2012} used a value of $-1/2$ in their studies. This value should be positive since the wind speed matches the Alfv\'en speed at the Alfv\'en radius and that our winds accelerates with distance to the star. In order to propose a value for $q$, we computed an average value of the Alfv\'en speed along the Alfv\'en surface, for slowly rotating cases ( $f \leq 10^{-2}$), since the rotational influence is already included through the parameter $K_2$ and $K_4$. We also focused our sample on the range where most of our Alfv\'en radii are, namely $R_A/R_* \in [2,13]$, which represents 26 out of our 60 simulations. Then, as shown in Figure \ref{AlfSpeed}, the variation of the Alfv\'en speed (or wind speed) with the Alfv\'en radius can be approximated with the following equation:

\begin{equation}
v(R_A) = v_{esc} (a(R_A/R_*)^q+b),
\label{vra_fit}
\end{equation}

with $q=0.7$, $a=0.063$ and $b=-0.084$.

The offset is due to the fact that the wind starts to accelerate at the surface of the star, \textit{i.e.} at $R/R_*=1$. Indeed, when magnetic field tends towards zero, the Alfv\'en surface is located just outside the surface where the wind speed start to increase. Interestingly the value of $m_{th}$ for the quadrupolar and the octupolar topologies (see Table \ref{threetoppar}) are very close to the fitted value obtained with our simulations. Moreover, in the case of a purely radial field (or a split monopole), we find that the value of $m_{th}$ is close to the value obtained with our topology independent formulation using the open flux, \textit{i.e.} the part of the flux created by the open/radial field lines. However, even if the value $q=0.7$ yields good fits for small Alfv\'en radii, it is too high to fit our Alfv\'en radii around $30 R_*$. Indeed, as shown in \citet{Parker1958}, the speed profile behaves asymptotically as the square root of a logarithm in the hydrodynamical case and $q$ diminishes with the distance to the star. This is why the $m_{th}$ value for the dipolar cases is less satisfactory and would need a lower value of $q$. Nevertheless any estimation should be positive and thus different than the one from \citet{Kawaler1988} and \citet{ReinersMohanty2012}.

\section{The case of realistic mixed topologies of the Sun and young stars}
\label{Realistic}

The formulation given in the previous section seems to work well on single-mode topologies. But what happens if we mix all three components in order to make more realistic simulations? The topology of the Sun changes during its cycle, from mostly quadrupolar during activity maximum to a mostly dipolar global topology at its minimum. This topology change has been measured by \citet{DeRosa2012} with data of the Wilcox Solar Observatory and the MDI instrument on board SoHO. This work gives the coefficients of each component of the spherical harmonics decomposition of the magnetic field normal to the solar photosphere. We used the classical formalism of this decomposition \citep{Donati2006}, \citep{DeRosa2012}, but using only the three axisymmetric components $l=1,2,3$ :

\begin{figure*}
\begin{tabular}{ccc}
\includegraphics[scale=0.7]{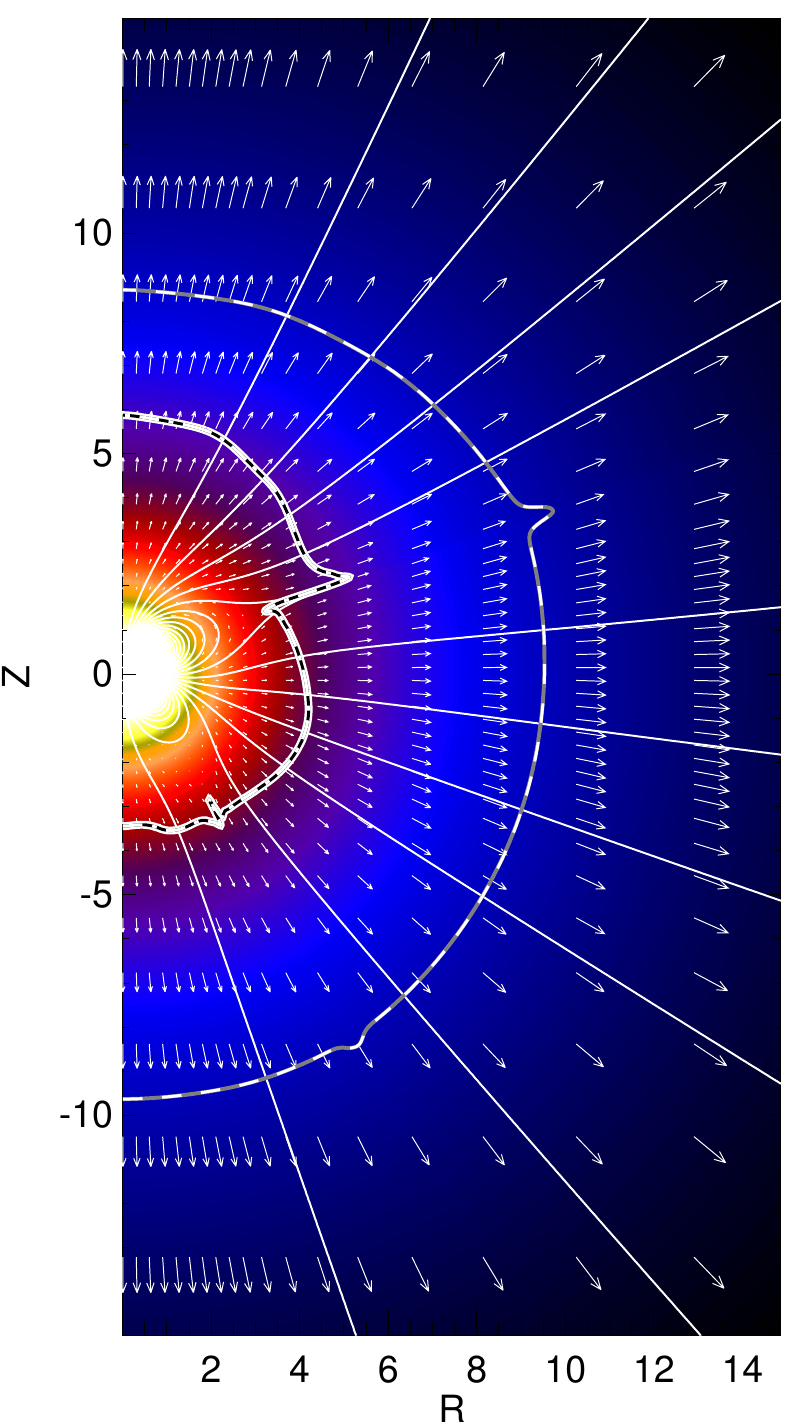} & \includegraphics[scale=0.7]{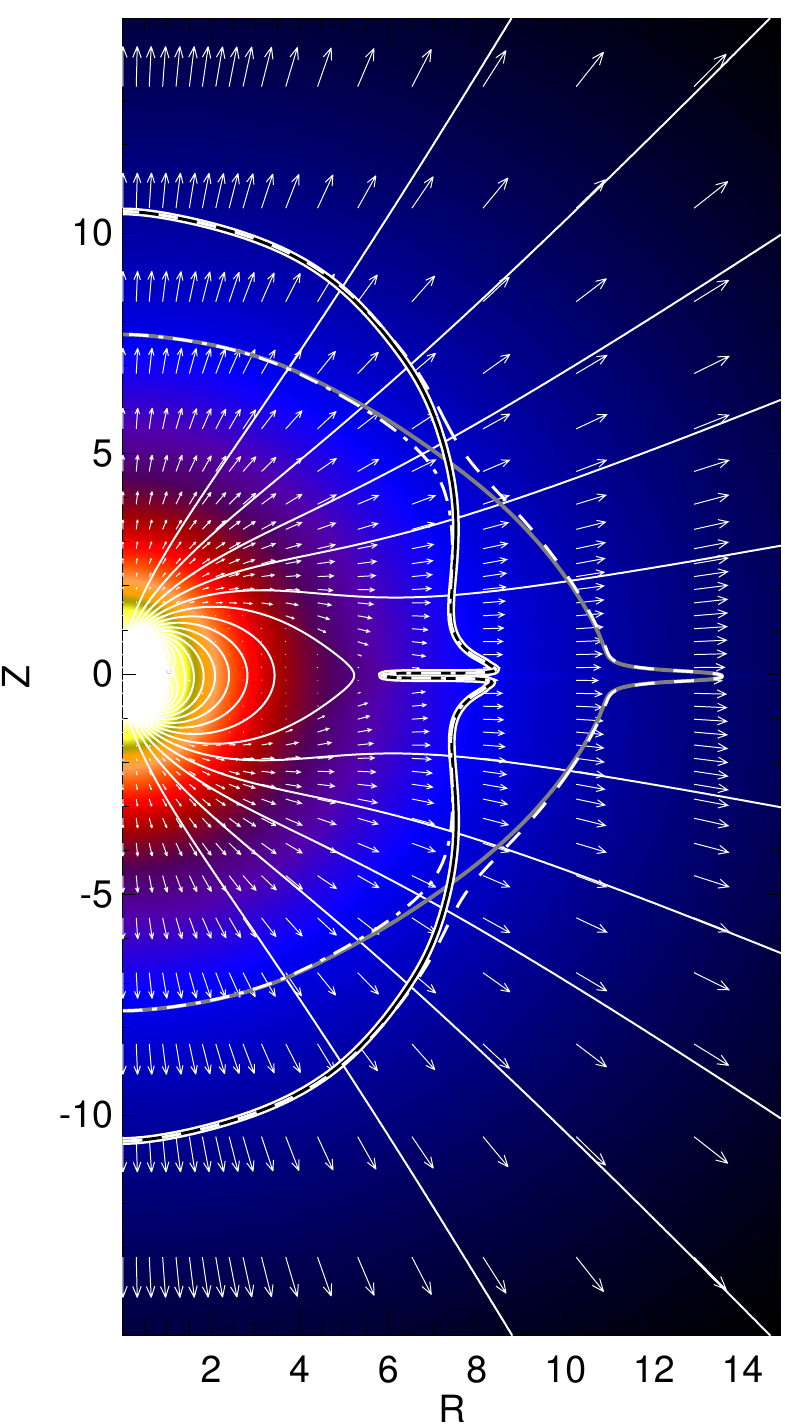} &\includegraphics[scale=0.75]{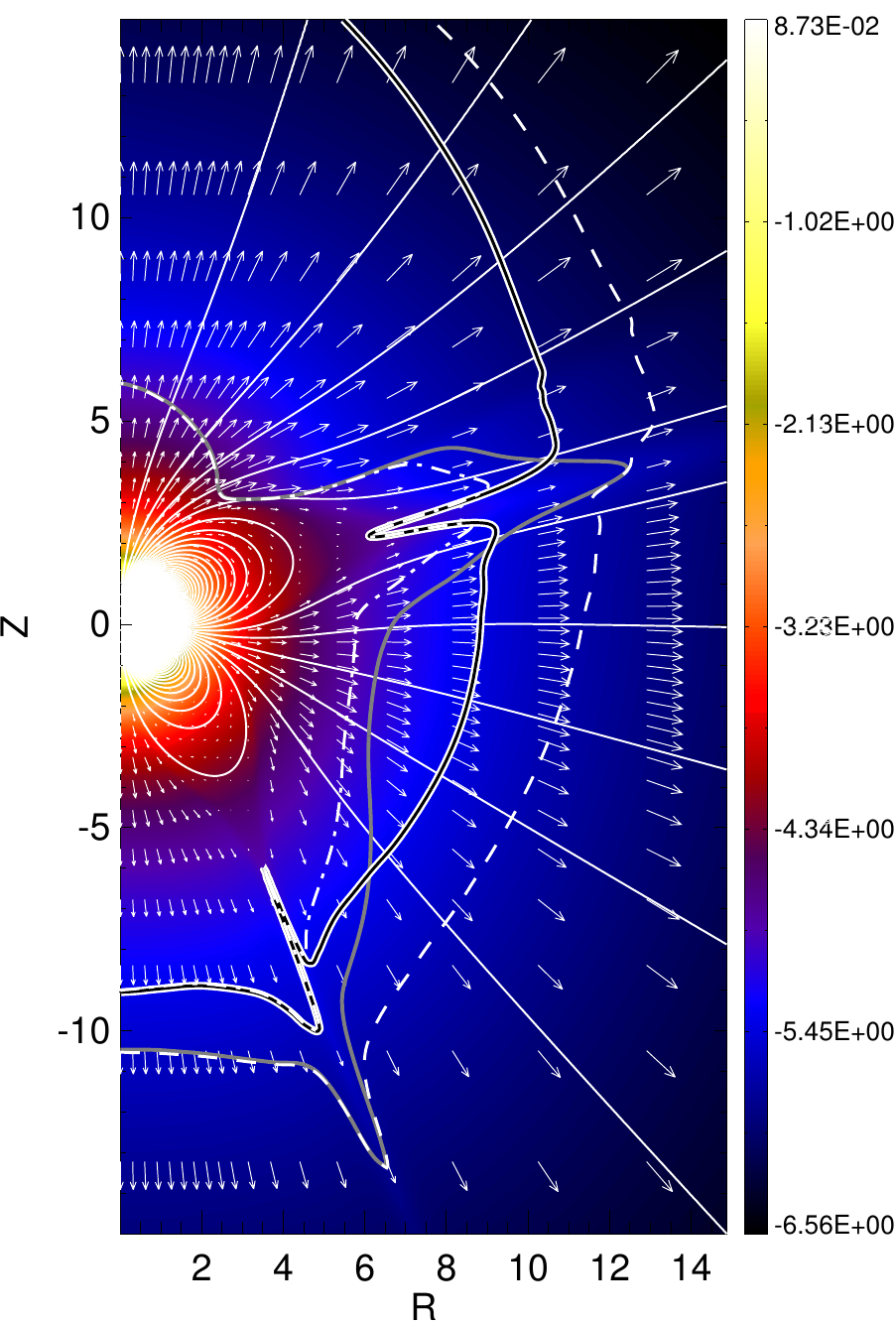} \\
\textbf{Solar Maximum} & \textbf{Solar Minimum}& \textbf{TYC-0486-4943-1}\\
\end{tabular}
\caption{Steady state solutions of three winds with a realistic magnetic topology extracted from Wilcox Solar Observatory data \citep{DeRosa2012} for the solar cases and from a ZDI Map of TYC-0486-4943-1. Only the axisymmetric component till $l=3$ are taken into account. The wind critical surfaces are shown in the same format as Figure \ref{WindVis}. The background is the logarithm of the density and we added velocity arrows in white.}
\label{RealSol}
\end{figure*}

\begin{equation}
B_r(\theta,\phi,t)=\sum_{l=1}^{l=3} B_l^0(t) Y_l^0(\theta,\phi)
\end{equation}

where

\begin{equation}
Y_l^0(\theta,\phi)=C_l^0 P_l(cos(\theta)),
\end{equation}

\begin{equation}
C_l^0=\sqrt{\frac{2l+1}{4\pi}},
\end{equation}

and $P_l^m$ are the Legendre polynomials. We performed two simulations with the coefficients given in Table \ref{RealPar}; the results for the Sun are given there as well. Adding those three components together creates a complex topology (see Figure \ref{RealSol}). In the case of the Solar maximum, the topology is close to quadrupolar, while at the minimum, a strong dipole dominates.

\begin{deluxetable}{lccc}
\tablecaption{Parameters and results of the realistic topology simulations.\label{RealPar}}
  \tablecolumns{4}
  \tabletypesize{\scriptsize}
  \tablehead{
    \colhead{} &
    \colhead{Sun Min} &
    \colhead{Sun Max} & 
    \colhead{TYC-0486}
  }
  \startdata
  Parameters & & &\\
  \hline
  $B^0_1$ (G)& -4.5 & 0.5 & -4.5 \\ 
  $B^0_2$ (G)& 0.24 & 2.2 & -28.2 \\
  $B^0_3$ (G)& -4.24 & 0.5 & -4.7 \\
  $f$ & 0.004 & 0.004 & 0.03 \\
  Period (d) & 28 & 28 & 3.75 \\
  \hline
  Results & & & \\
  \hline
  $\Upsilon_{open}$ & 155 & 14 & 275 \\
  $R_A/R_*$ & 6.7 & 3.3 & 7.7 \\
  $\dot{M}_w$ ($10^{-14}$ $M_{\odot}$/yr) &0.85 & 0.87 & 0.9 \\
  \enddata
  \tablecomments{Input parameters are given in physical units here assuming a density normalization $\rho_0=2.9 \time 10^{-15}$ g/cm$^3$ and a magnetic field normalization $B_0=8.3$ G. The agreement of the results of those simulations and the topology independent formulation is shown in Figure \ref{Col3par}.}
\end{deluxetable}

In order to test our stellar wind model and topology-independent formulation on another star than the Sun, we used output from Zeeman-Doppler Imaging (ZDI) Maps \citep{Donati2006} of young stars (observed with NARVAL at T\'elescope Bernard Lyot) with a strong non-dipolar magnetic field. As young stars are generally more magnetic than the Sun, we expect a larger Alfv\'en radius, although this depends strongly on the unknown mass loss rate. The case of the 70 million year old star TYC-0486-4943-1 is interesting because it has a dominant quadrupolar component and a strong magnetic activity. The radius and the mass of this star are $R_{TYC}=0.68 R_{\odot}$ and $M_{TYC}=0.69 M_{\odot}$, so the speed normalization can remain the same since the keplerian speed ($v_{kep}=\sqrt{GM_*/R_*}$) is very similar to the Sun. Its rotation period is $3.75$ days and thus falls within the range of our parameter study ($f=0.03$). The coronal temperature ($T_c$) may be higher \citep{Preibisch1997} in this star, since it is more active magnetically, but without more information we chose to keep the same $c_s/v_{esc}$ for all simulations. Changing the value of the coronal temperature is likely to change the values of the constants $K_1$ and $K_3$ that would diminish with higher $T_c$ \citep{MP2008,Matt2012,UdDoula2009}, and the values of $K_2$ and $K_4$. We expect the exponent $m$ to be robust to this change but a more systematic study is needed, which is beyond the scope of this paper.

The winds created by all those configurations are shown in Figure \ref{RealSol}, and the results are listed in Table \ref{RealPar}. We see in Figure \ref{Col3par} that the torque of those wind are well described by our formulation. We tested other configurations of topologies and all fall onto our law (not shown here). Interestingly, despite a much higher magnetic energy density for the young K-star TYC-0486 ($\propto \sum (B^l_m)^2$), the average Alfv\'en radius is only slightly larger than for the Solar minimum. This results demonstrates how relevant is the topology parameter for the calculation of the magnetic torque. It is also interesting to note that during one cycle the magnetic braking (which is proportional to $R_A^2$) vary by a factor 4 (\citealt{PintoBrun2011} found a factor 16 over the solar cycle, see also \citealt{Vidotto2012}). This could have an effect on the long time rotational evolution of the Sun.

 The value for the Alfv\'en radius in both solar cases is low compared to the expected value of $10-12 R_{\odot}$ \citep{Pizzo1983}. But depending on models the range can vary between $2.5 R_{\odot}$ and $60 R_{\odot}$. Isothermal models, from which we are close tend to to give a lower limit for $R_A$ \citep{PneumanKopp1971}, whereas conductive models give the highest estimates \citep{DurneyPneuman1975}. The mass loss rates are also lower than the usual value for the Sun ($2-4.10^{-14} M_{\odot}$/yr), and since we do not know the mass loss rate of TYC-0486-4943-1 we kept the same density normalization $\rho_0= 2.9\times 10^{-15}$g/cm$^3$ for all the values of mass loss rates given in Table \ref{RealPar}. We will come back to this point in section \ref{Mdot}.

\section{Discussion}
\label{discuss}

\subsection{Mass loss rate}
\label{Mdot}

In section \ref{Realistic} we give values for the solar mass loss rate. Here normalization plays an important role. We chose the density normalization to be $\rho_0=2.9\times10^{-15}$ g/cm$^{3}$. The mass loss rate of the two cases is then around $0.8 \times 10^{-14} M_{\odot}/$yr. We could simply change the density normalization to get solar mass loss rates, but we would then have lower values of the Alfv\'en radii in our Solar cases, given that the magnetic field normalization is imposed by the density normalization.

However the mass loss rate is also very sensitive to the parameter we kept fixed: $c_s/v_{esc}$ and $\gamma$. For instance increasing temperature by 12\% to get $c_s/v_{esc}=0.235$  multiplies the mass loss rate by 5. Thus there is some freedom that can be used to get closer to solar values, changing temperature and normalizations.

Also the mass loss rate depends on physics not included in our simulations. For instance our simulations include no heating, and the driving could be better physically modeled including Alfv\'en waves. This is why we give our braking laws as a function of the mass loss rate, so that rotation evolution model can take into account much more physics than we do here, for instance using the method proposed by \citet{CranmerSaar2011}.

\begin{figure}
\includegraphics[scale=0.55]{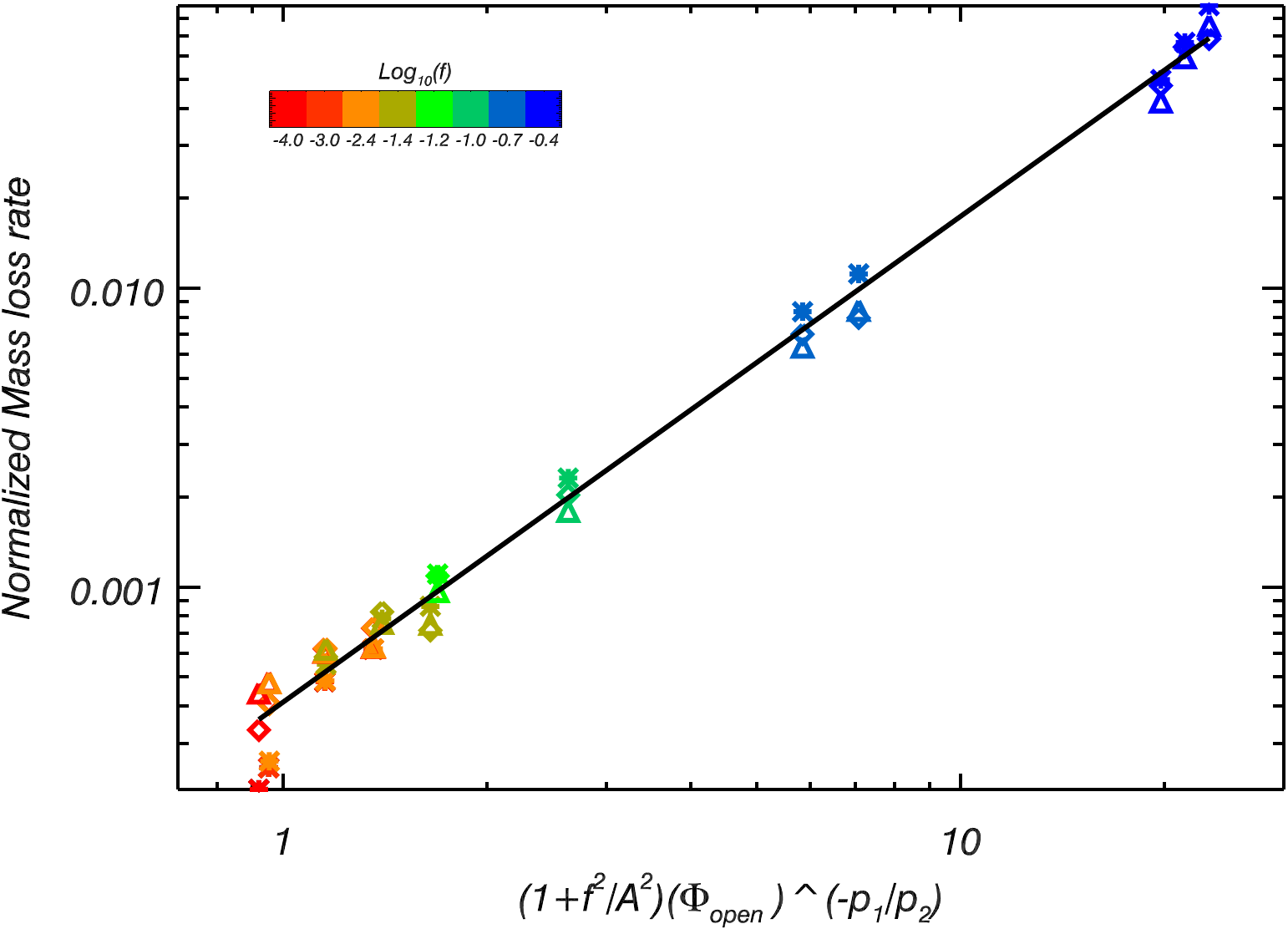}
\caption{Fit of $\dot{M}_w$ with the stellar parameters $f$ and $v_A/v_{esc}$ (see equation \ref{mdotform}).}
\label{mdotfit}
\end{figure}

Nevertheless, we performed a simple fit of the mass loss rate in the case of our coronal temperature, as a function of the parameters we varied: $v_A/v_{esc}$ and $v_{rot}/v_{esc}$. As $\dot{M}_w$ is an increasing function of $f$ and a decreasing function of $v_A/v_{esc}$ we propose the following formulation:

\begin{eqnarray}
\dot{M}_w &=A_1 (v_A/v_{esc})^{-p_1}(1+\frac{f^2}{A_2^2})^{p_2}\\
&= A_1 (B_* \sqrt{\frac{R_*}{8 \pi \rho_* GM_*}})^{-p_1}(1+\frac{f^2}{A_2^2})^{p_2}
\label{mdotform}
\end{eqnarray}

fitting this formula with our set of simulations we find: $A_1=0.28$, $p_1=0.19$, $A_2=0.087$ and $p_2=1.6$. It means that $\dot{M}_w$ is strongly increased at high rotation while the dependence on the magnetic field strength is rather weak (there is a factor 8 between $p_1$ and $p_2$). This fit is shown in Figure \ref{mdotfit}. 

\citet{Cohen2014} found $\dot{M}_w \propto B_*$, which seems contradictory with our results. However their wind driving is related to an energy source term that changes accordingly with the magnetic field. As we see with the value of our fit, the dependence of $\dot{M}_w$ on the magnetic field strength is weak, whereas the coronal temperature (which we held fixed) has a very strong influence on $\rho v$. Hence a trend such as found in \citet{Cohen2014} is expected, because the coronal heating is a consequence of the magnetic activity. But in our parameter study with a fixed temperature, polytropic wind, a higher strength of the magnetic field reduces the mass-loss rate, since more magnetic loops are formed and confine the plasma.

\subsection{Higher order components of the magnetic field and non-axisymmetry}

We have limited our study to three spherical harmonics modes. ZDI maps offer many more non-zeros components of star's magnetic fields and our formulation could be further generalized. For axisymmetric modes our formulation should be robust. Indeed, as we go to higher orders, the radial decay of the magnetic field increases and we can already see that the octupole has very little influence on the large scale topology of the steady state solution of the wind when mixed with a dipolar or a quadrupolar field (if the surface amplitudes are comparable). In any case, the open flux captures the effect of higher order fields by decreasing if more plasma is kept in magnetic loops. 

However as all our study is axisymmetric, a full 3D study of the torque created by complex topologies must be investigated in order to fully test our formulation.

\section{Conclusion}

In this work we give the first quantitative results on the systematic influence of the topology of stellar magnetic fields on the Alfv\'en radius and the torque applied by a magnetized wind on its star. Our formulation in the simplest dipolar case is very similar to \citet{Matt2012}. We found that the more complex the field is the smaller is the torque. Qualitatively this has been expected from simple scaling arguments \citep[see quadrupolar simulations in][]{MP2008} and we are now able to give braking laws in three ideal axisymmetric cases: a dipole, a quadrupole and an octupole. Furthermore we derived a unified topology-independent formulation from our set of 60 simulations. In the case of our Sun today, the first two topologies are dominant during the activity cycle. We have performed simulations of the two cases of the solar minimum and maximum, whose torque is well predicted by our formulation. In the cases of more active stars, like the young K-star TYC-0486-4943-1, even higher order magnetic multipoles can be significant during an activity cycle, and our topology-independent formulation is a first step to understand the influence of a fully realistic topology on the angular momentum loss. With this formulation the equivalent torque is given as following:

\begin{eqnarray}
\tau_w&=\dot{M}_w \Omega_* R_*^2 K_3^2 (\frac{\Upsilon_{open}}{(1+f^2/K_4^2)^{1/2}})^{2m}\\
& = \dot{M}_w^{1-2m} \Omega_* R_*^{2-4m} K_3^2 (\frac{\Phi_{open}^2}{v_{esc}(1+f^2/K_4^2)^{1/2}})^{2m}.
\end{eqnarray}

However, considering the present day knowledge of the mass loss rate and the wind velocities at 1 AU, our model might need to include more physics or at least an exploration of the other parameters, \textit{i.e.} $\gamma$ and $c_s/v_{esc}$.  Also, even if the open flux is more difficult to get than a star's magnetic field strength at the equator from observations, our study shows that it might more meaningful for torque calculations. Further works could tackle the prediction of open flux from knowledge of surface field from ZDI maps (as in \citealt{Vidotto2014}) and provide torque for distant stars.

A major step to further confirm this formulation is to test it on non-axisymmetric realistic topologies, and this is left for further works.

\section{Acknowledgements}
We would like to thank Colin Folsom, Pascal Petit for the magnetic field decomposition coefficients of TYC-0486-4943-1, J\'erome Bouvier and the ANR TOUPIES project which aim to understand the evolution of star's spin rates, the ERC STARS2 (www.stars2.eu) and CNES support via our Solar Orbiter funding.

\bibliographystyle{yahapj}
\bibliography{biblio}

\begin{appendix}

\section{Derivation of torque formulations}
\label{AppA}

In order to have a semi-analytic expression for the Alfv\'en radius thus a torque formulation, a one dimensional approach has been introduced by \citet{Kawaler1988}. Assuming that the magnetic field strength decreases as a single power law $B=B_*(R_*/R)^{l+2}$ we have at the Alfv\'en radius the following equality:
\begin{equation}
v(R_A)^2 = v_A^2 = \frac{B^2}{4 \pi \rho_{R_A}} = \frac{B_*^2 R_*^{2l+4}}{R_A^{2l+4} 4 \pi \rho_{R_A}},
\end{equation}

and thus: 

\begin{equation}
(\frac{R_A}{R_*})^{2l+2} = \frac{B_*^2R_*^2}{4\pi R_A^2 \rho_{R_A} v(R_A) v(R_A)}.
\end{equation}

Interpreting $4 \pi \rho_{R_A} v(R_A) R_A^2$ as the mass loss rate at this radius and considering a simple model for the dependence of the Alfv\'en speed at the Alfv\'en radius:

\begin{equation}
v(R_A) \propto v_{esc} (R_A/R_*)^q,
\label{vra}
\end{equation}

we are left with:

\begin{equation}
(\frac{R_A}{R_*})^{2l+2+q} \propto \frac{B_*^2R_*^2}{\dot{M}_w v_{esc}} \equiv \Upsilon.
\label{Alf1}
 \end{equation}

The parameter $q$ is likely positive (the wind accelerates) even though in the literature it has often been set to $-1/2$ (\citealt{Kawaler1988,ReinersMohanty2012}). Hence we have a semi-analytical formulation for the Alfv\'en radius:

\begin{equation}
\frac{R_A}{R_*} = K \Upsilon^m,
\label{simple_power_law}
\end{equation} 

where

\begin{equation}
m=1/(2l+2+q).
\label{slope}
\end{equation}

Although in two dimensions the Alfv\'en surface is not spherically symmetric and the magnetic field does not follow a single power law, \citet{MP2008} demonstrated that the formulation (\ref{simple_power_law}) fits the simulations results precisely when all the constituting parameters of $\Upsilon$ ($B_*$, $v_{esc}$, $\dot{M}_w$) vary.

In \citet{Matt2012}, 50 simulations were performed allowing the magnetic field strength and the rotation rate to vary. Considering that the rotation rate mainly appears in the magneto-centrifugal effect, one can write a modified speed:

\begin{equation}
v_{mod}^2 = v_{esc}^2 + \frac{2\Omega_*^2 R_*^2}{K_2^2} = v_{esc}^2(1+f^2/K_2^2),
\end{equation}
where $f$ is the fraction of break-up rate defined in section \ref{StellarWindModels}.

The formulation for the Alfv\'en radius as a function of $(\Omega_*,\Upsilon)$ is given by replacing $v_{esc}$ in equations (\ref{vra}) and (\ref{Alf1}) by this modified speed $v_{mod}$:

\begin{equation}
\frac{R_A}{R_*}=K_1 [\frac{\Upsilon}{(1+f^2/K_2^2)^{1/2}}]^m.
\end{equation}

This formulation precisely captures the influence of the rotation rate of the star on the magnetic torque in the 50 simulations of \citet{Matt2012}\footnote{The formulation \ref{form1} is equivalent to but algebraically modified compared to \citet{Matt2012} so that coefficients $K_1$ and $K_2$ are more meaningful.}. The values found for the three models parameters are: 

\begin{equation}
K_1 = 2.5, \quad K_2 = 0.07, \quad m = 0.22.
\end{equation}

The parameter $K_2$ is the value from which rotation rate starts to have a influence on the magnetic braking. Indeed the magneto-centrifugal effect adds acceleration to the wind and thus the Alfv\'en radius gets closer to the star. Rapid rotation also can increase the mass loss rate (for a given $v_A/v_{esc}$) but this effect is already taken care of in the parameter $\Upsilon$.

\section{Boundary conditions and conservation properties}
\label{AppB}

\begin{wrapfigure}{R}{0.5\textwidth}
\center
\includegraphics[scale=0.6]{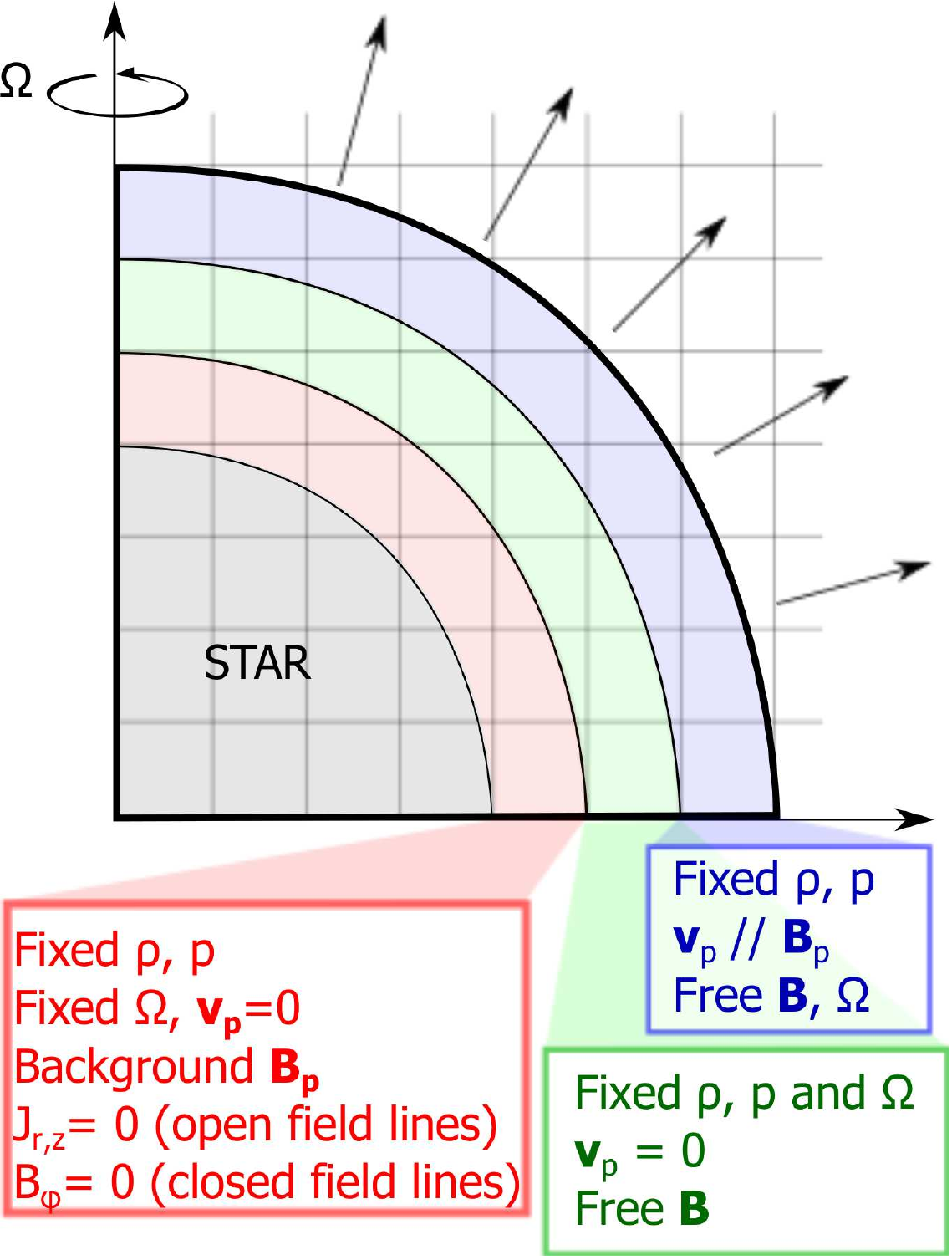}
\caption{A three layer Boundary Conditions is used to ensure the conservation of quantities such as $\Omega_{eff}$, which measure the effective rotation of the magnetic field lines with the star.}
\label{BC}
\end{wrapfigure}

In this work the boundary conditions at the external limits of the domain are outflow conditions except for the axisymmetric axis\footnote{Outflow boundary condition set zero gradient across the boundary for all variables while axisymmetric condition along the rotation axis is such that variables are symmetrized across the boundary and the normal and toroidal components of vector fields flip signs.}. The star is then modeled through a three layer boundary condition described in Figure \ref{BC} \citep{Strugarek2014IAU}. Each layer is at least one grid cell thick. The interior of the star is not modeled so that quantities into the grey area on Figure \ref{BC} are set to constant values. In the red layer the poloidal field is set to be as in the initial state (dipolar, quadrupolar, octupolar or a mix of the three) and since we use a background field splitting, we set the fluctuating field to zero. The poloidal speed is set to zero, while the star is in solid body rotation. The density and pressure profile are fixed to the polytropic solution of Parker's equations for a one dimensional hydrodynamical wind with $\gamma=1.05$ \citep[see polytropic solutions of][]{KG2000,LC1999}. A sophisticated condition is used on the toroidal field, which is set to maintain the poloidal current as close to zero as possible. This is an empirical method first used in \citet{MB2004}, that allow good conservation properties of the effective rotation, which is key for torque calculations.

However this condition is only necessary where the field lines are open and we set the toroidal field to zero on closed field lines regions to avoid artificial creation of $B_{\phi}$ in the dead zones. We dynamically discriminate open/close field lines regions using an empirical criterion on the Alfv\'en speed associated with the azimuthal magnetic field in the upper layers of the boundary conditions.

In the green layer, the same conditions are applied to the pressure, density, rotation rate, and the poloidal speed is set again to zero, but we let the magnetic field evolve. In the blue layer we keep the same conditions for $\rho$ and $p$ while the poloidal speed and magnetic field are forced to be parallel.

We give the analytic expressions of the magnetic field used to initialize our simulations, for the dipole, the quadrupole, or the octupole in cylindrical coordinates, where $B_*$ is the amplitude at the equator:\\

\textbf{Dipole}
\begin{equation}
B_{R} = B_*\frac{3 RZ}{r^{5}} ,\quad B_{Z} = B_*\frac{( 2 Z^{2} - R^{2})}{r^{5}} 
\end{equation}

\textbf{Quadrupole}
\begin{equation}
 B_{R} = B_*\frac{R (5 (Z/r)^{2}-1)}{r^{5}},\quad B_{Z} = B_*\frac{ Z(5 (Z/r)^{2}-3)}{r^{5}}
\end{equation}

\textbf{Octupole}
\begin{equation}
B_{R} = B_*\frac{(35 Z^{3}R/r^{2}-15ZR)}{3r^{7}}, \quad B_{Z} = B_*\frac{(35 (Z/r)^{4}-30(Z/r)^{2}+3)}{3r^{5}}
\end{equation}

\begin{figure}[h!]
\center
\includegraphics[scale=0.12]{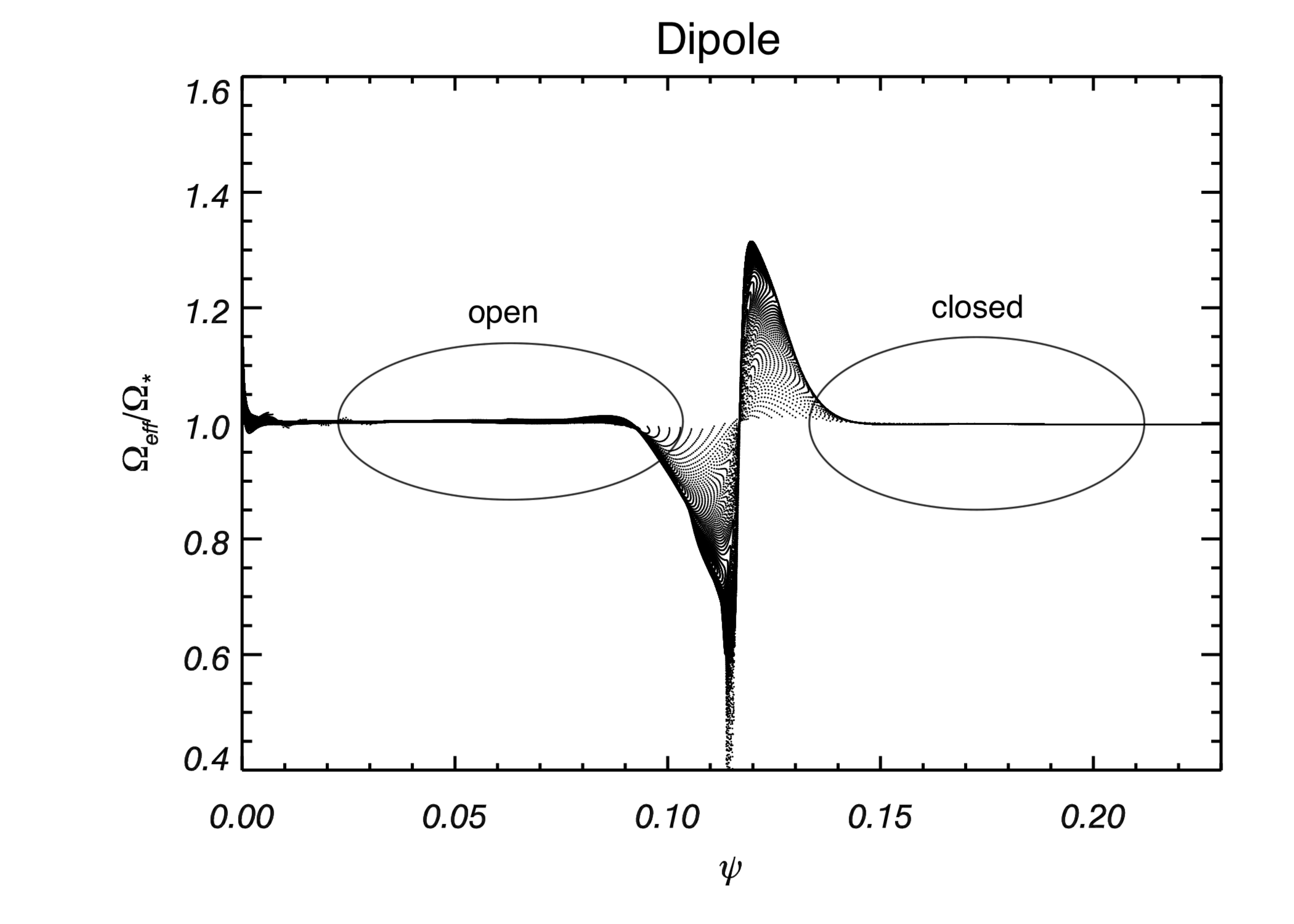}
\includegraphics[scale=0.12]{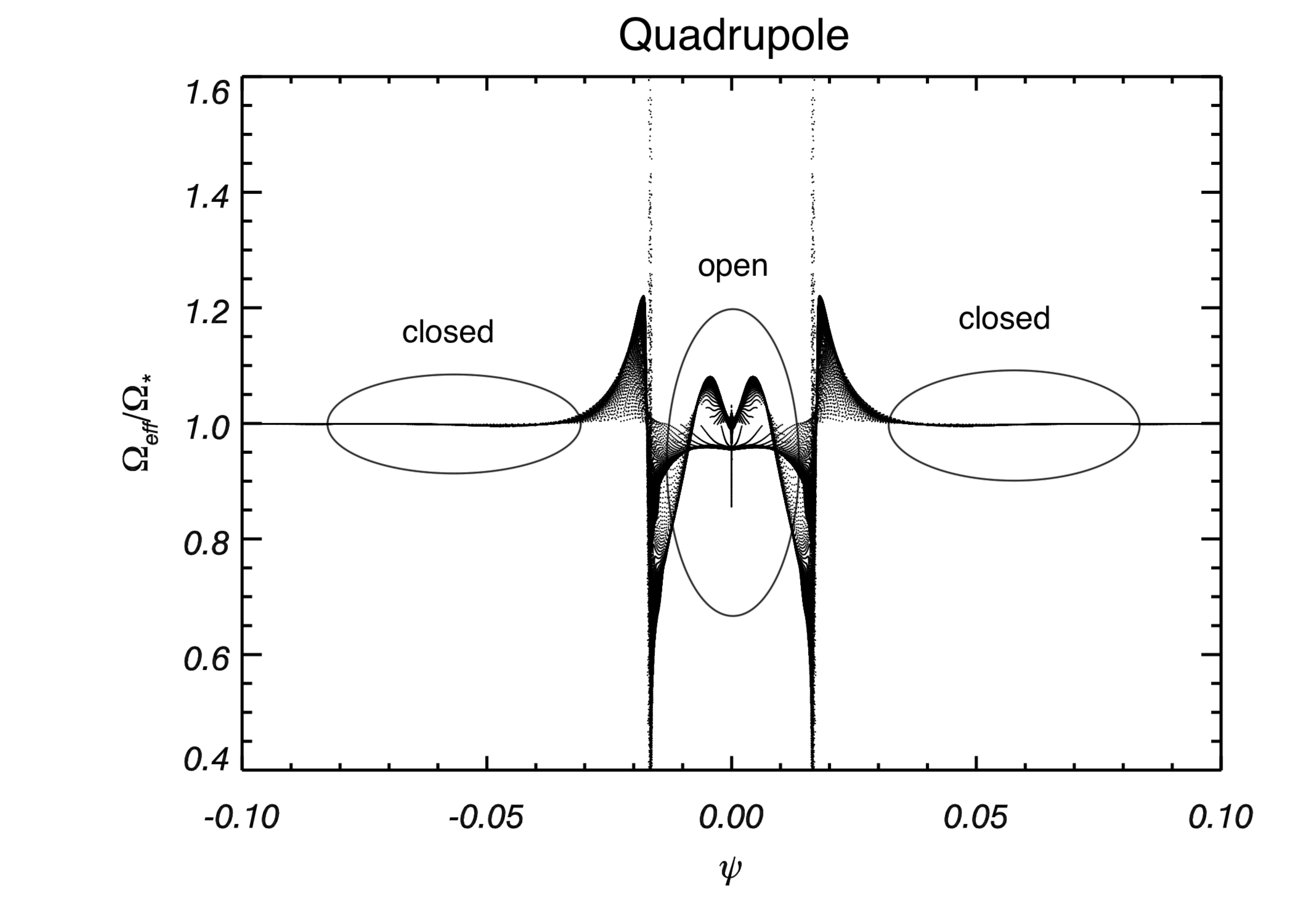}
\includegraphics[scale=0.12]{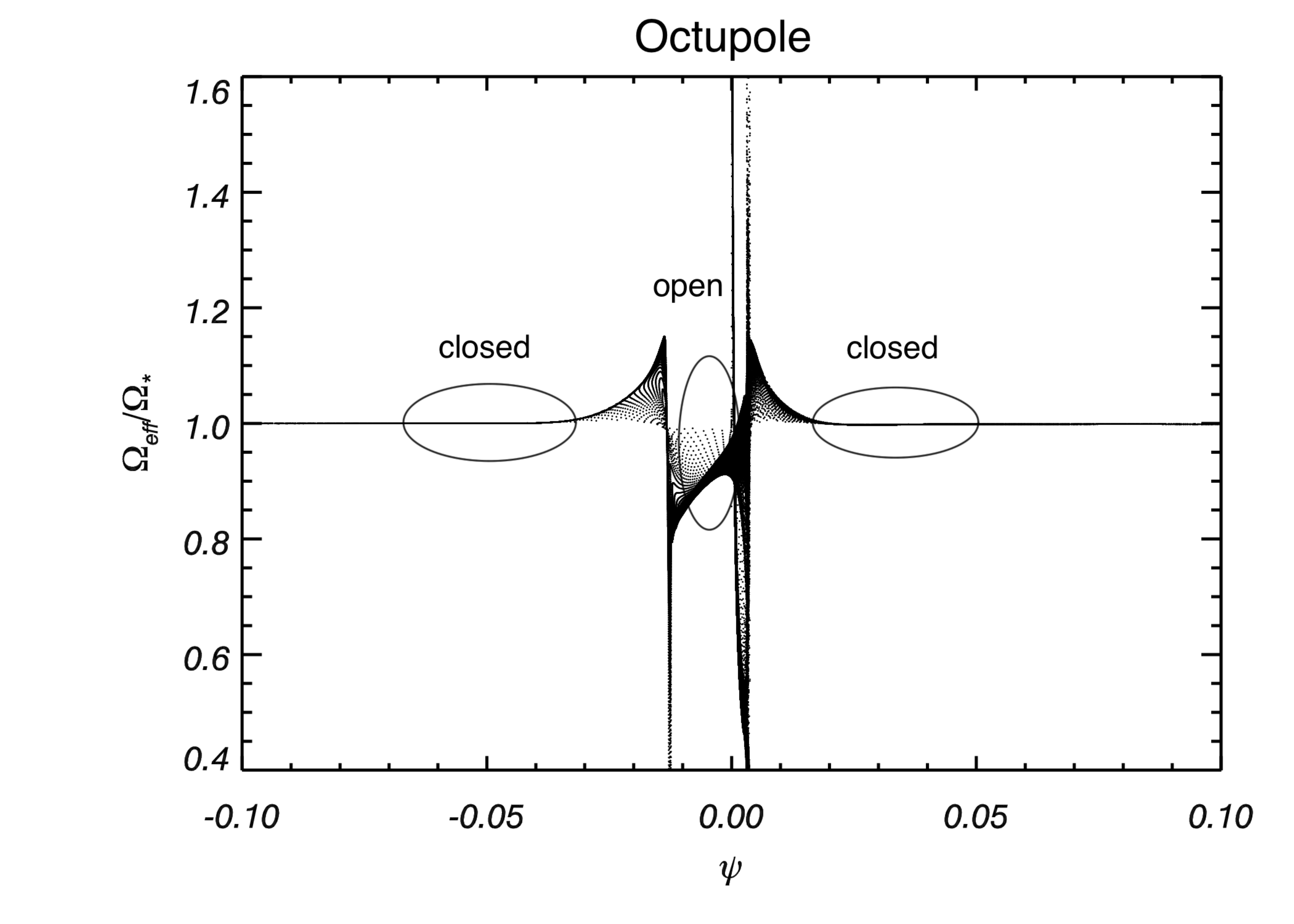}
\caption{Effective rotation for the three topologies in case 31: dipole on the top left, quadrupole on the top right and octupole on the bottom.}
\label{roteff}
\end{figure}

Doing variational principle with the Lagrangian of ideal MHD equations in the axisymmetric case, it can be demonstrated that some scalar quantities must me be conserved along the magnetic field lines. Among them is the energy (the Bernoulli function), the entropy and the effective rotation rate of the magnetic field lines, which is the derivative of the electric field potential \citep{KG2000,ZanniFerreira2009}, defined as follows:
\begin{equation}
\Omega_{eff}(\psi) \equiv \frac{1}{R} (v_{\phi}-\frac{v_p}{B_p}B_{\phi})
\end{equation}

where the subscript $p$ stands for the poloidal component of the field, and the subscript $\phi$ for the toroidal component, thus we have $B_p=\sqrt{B_R^2+B_Z^2}$. $\Omega_{eff}$ is expressed as a function of $\psi$, the stream function of the magnetic field which can be computed as $\psi=RA$, where $R$ is the cylindrical radius and $A$ the scalar potential of the magnetic field ($\mathbf{B}= \nabla \times (A \mathbf{e_{\phi}})$). As a consequence, since each value of $\psi$ can be associated with a magnetic field line, the conservation of this quantity is visualized on the full 2D grid through little vertical spread in the $\Omega_{eff}/\Omega_*$ versus $\psi$ plots in Figure \ref{roteff}. 
 
In steady-state ideal MHD, $\Omega_{eff}$ should be the same on all the magnetic field lines and equal to the rotation rate of the star in order to behave as field lines anchored into rotating stellar surface. Thus good conservation properties of $\Omega_{eff}$ are necessary to compute accurate value of the angular momentum flux. The conservation is affected by boundary conditions. For instance, if our deepest layer boundary condition on $B_{\phi}$ is not applied, the value of $\Omega_{eff}$ is below one by more than 20\% on the left arm on Figure \ref{roteff} for the dipolar case which corresponds to open field lines \citep[see][]{Strugarek2014IAU}. For other topologies, the value of $\Omega_{eff}/\Omega_*$ for open field lines remain close to unity even if non ideal features (current sheets) have more influence than in the dipolar case. Those non-ideal features are due to current sheet around the streamers, created by numerical diffusion. However using our resolution they have negligible influence on our average value of $R_A$ (this has been checked by increasing resolution, while those features diminish, our integrated values remain the same within 1\%).

\newpage
\section{Exhaustive Results}
\label{AppC}

\begin{figure}[h!]
\center
\includegraphics[scale=1]{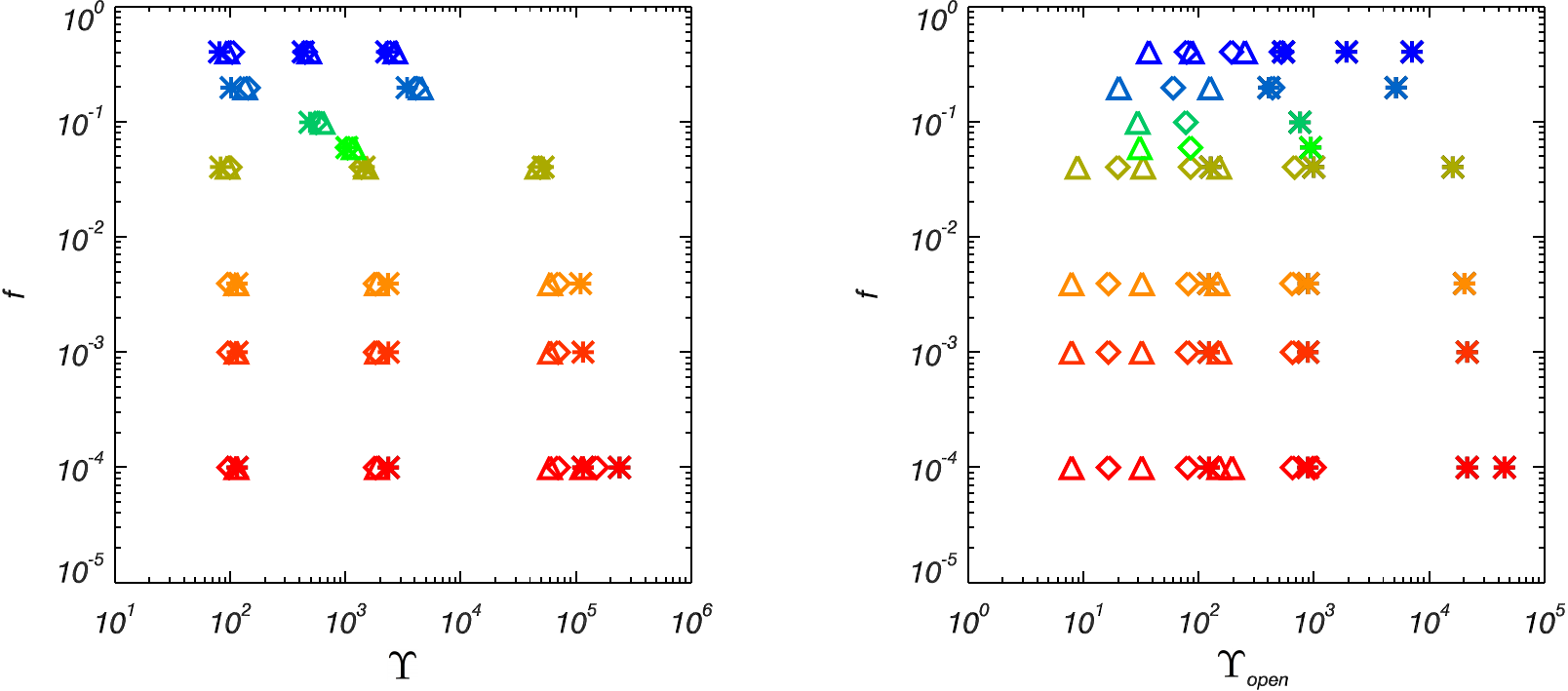}
\caption{Parameter space explored for the three different topologies: crosses are for the dipole, diamonds for the quadrupole and triangles for the octupole. As in Figures \ref{threetop}, \ref{Col2par}, \ref{Col3par} and \ref{mdotfit} colors are associated with rotation.}
\label{parspace}
\end{figure}

\begin{deluxetable}{lrrrrrrrrr}[h!]
\tablecaption{Results of our sixty simulations.\label{restable}}
  \tablecolumns{10}
  \tabletypesize{\scriptsize}
  \tablehead{
    \colhead{} &
    \colhead{} &
    \colhead{$\langle R_A \rangle$}&
    \colhead{} &
    \colhead{} &
    \colhead{$\Upsilon$} &
    \colhead{} &
    \colhead{} &
    \colhead{$\Upsilon_{open}$} &
    \colhead{} 
  }
  \startdata
  Case & dipole & quadrupole & octupole & dipole & quadrupole & octupole & dipole & quadrupole & octupole \\
  \hline
  &&&&&&&&&\\
  1 & 6.3 & 3.6 &	3.0&	115&	98&	113&	123&	16.5&	8\\
  2 &12.5&	5.3&	4.0&	2360&	1830&	1880&	886&	80&	32\\
  3 & 32.3&	9.3&	5.9&	115000&	70000&	59600&	21400&	652&	151\\
  3+ & 36.4&	9.9&	6.3&	239000	&151000	&114000	&50600	&1140	&199\\
  5 & 6.3 & 3.6 &	3.0&	115&	98&	113&	123&	16.5&	8\\
  6 &12.5&	5.3&	4.0&	2360&	1830&	1880&	886&	80&	32\\
  7 & 32.3&	9.3&	5.9&	115000&	70000&	59600&	20300&	646&	151\\
  8 & 6.3&	3.6&	3.0&	115&	98&	113&	122&	16.5&	8\\
  10 & 12.6 &	5.3&	4.0&	2340&	1830&	1875&	890&	81&	32\\
  13 & 32.3&	9.3&	5.9&	101000&	70600&	59900&	2030&	646&	145\\
  23 &5.9&	3.5&	3.1&	83&	99.5&	95&	127&	20&	9\\
  24 & 11.7&	5.2&	4.1&	1450&	1380&	1500&	998&	85&	33\\
  25 &30.3&	9.2&	5.8&	52000&	49000&	46200&	15000&	680&	151\\
  31 &10.6&	4.9&	4.0&	1030&	1045&	1180&	937&	85&	31\\
  37 & 8.7&	4.4&	3.6&	492&	560&	635&	758&	77.5&	30\\
  45 &5.5&	3.4&	2.9&	102&	143&	135&	407&	60&	20\\
  47 &13.4&	5.7&	4.4&	3430&	4085&	4510&	5170&	438&	126\\
  48 &4.9&	3.1&	2.6&	81&	104&	94&	545&	79&	37\\
  49 &7.1&	3.7&	3.2&	431&	447&	486&	1920&	193&	87\\
  50 &11.4&	5.0&	4.3&	2280&	2390&	2710&	7100&	524&	252\\
  \enddata
  \tablecomments{We give here all the computed values of $R_A$, $\Upsilon$ and $\Upsilon_{open}$ that were used to perform the fits of this paper. The parameters used for each cases are listed in Table \ref{inpartable}.}
\end{deluxetable}

Table \ref{restable} gives all the results that we used to fit our formulations. Figure \ref{parspace} shows the parameter space explored in terms of $\Upsilon$ and $\Upsilon_{open}$ and also visualizes the dependence of these quantities on the rotation rate. For instance we can see that $\Upsilon$ depends weakly on the topology (the different symbols are merged), this is not the case of $\Upsilon_{open}$, for which all three topologies are well separated for a given case. This is necessary to get a topology-independent formulation.

\begin{figure}
\center
\includegraphics[scale=0.55]{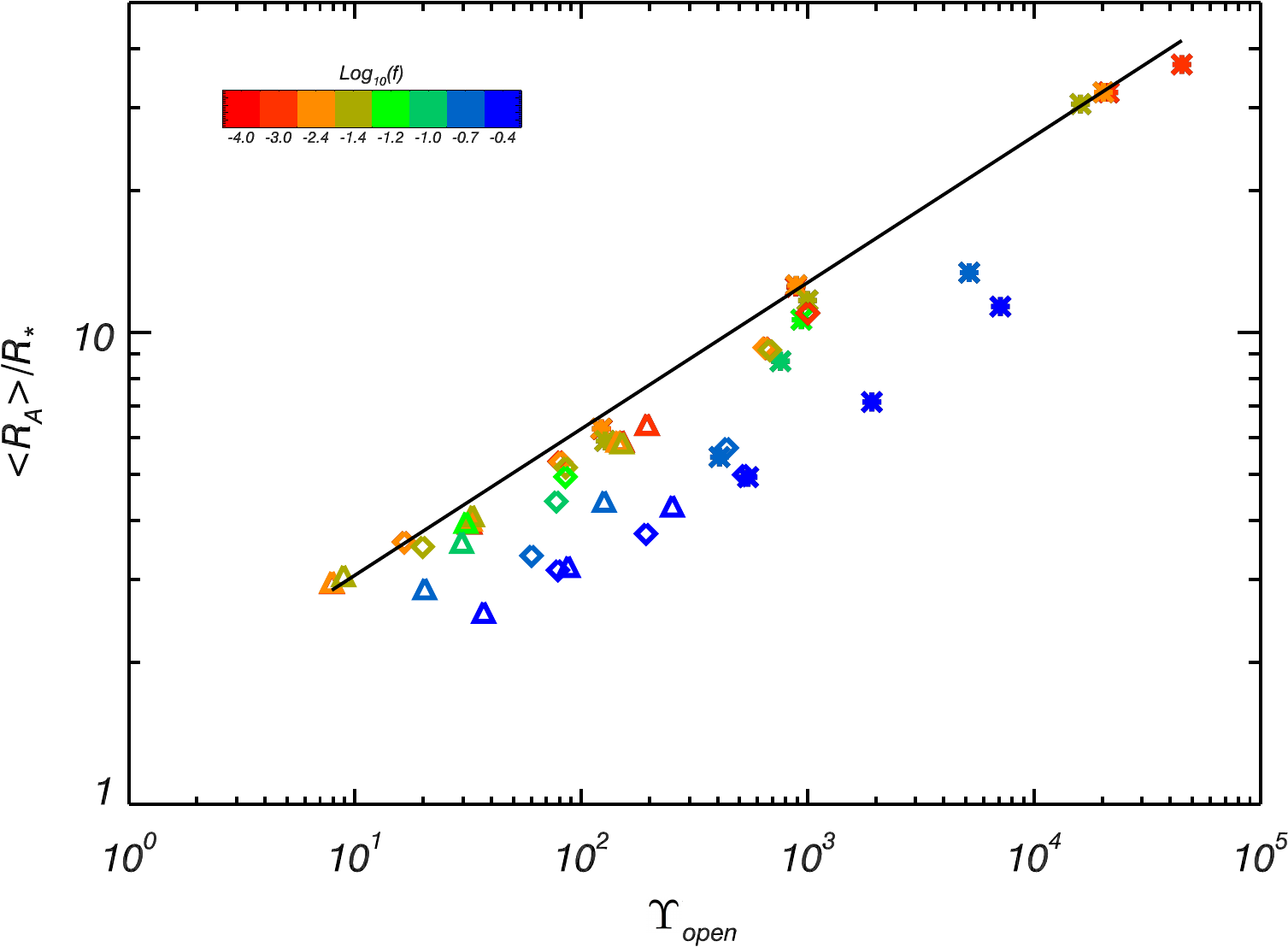}
\caption{Alfv\'en radius versus $\Upsilon_{open}$. Colors and symbols are the same as Figure \ref{threetop}. A constant slope (represented by the black line) is observed between simulations at the same rotation rate, but higher rotators (green, light blue, blue) are shifted to smaller $R_A/R_*$.}
\label{Col2par}
\end{figure}

Eventually, we would like to show how important it is to take into account the rotation rate in our formulation. For instance considering a simple formulation such as

\begin{equation}
\frac{R_A}{R_*}=K_1\Upsilon_{open}^m,
\label{Upsop_pow}
\end{equation}

 gives a general trend (see Figure \ref{Col2par}). For a given rotation rate, points of all topologies follow the same power-law. However for rotation rates beyond $f=0.05$ (green, light blue and blue points) the power-law is shifted downward.

\end{appendix}

\end{document}